\documentclass{article}
\usepackage{aaspp4}

\lefthead{Newbury \& Fahlman}
\righthead{Strong Gravitational Lens Reconstruction}

\begin{document}

\title{Reconstructions of the Strong Gravitational Lenses\\
  MS~2137 and MS~1455 Using a Two-Stage Inversion Algorithm}

\author{P. Newbury}
\affil{Department of Mathematics and Statistics, Langara College\\
  Vancouver, Canada, V5Y 2Z6\\
  pnewbury@langara.bc.ca}

\and

\author{G. G. Fahlman}
\affil{Department of Physics and Astronomy, University of British
  Columbia\\ Vancouver, Canada, V6T 1Z2\\
  fahlman@astro.ubc.ca}

\begin{abstract}
  We propose a new, two-stage algorithm for inverting strong
  gravitational lenses.  The key to the algorithm is decoupling the
  effects of lens magnification and intrinsic structure in the
  background source in the appearance of lensed arcs.  First, the
  distribution of mass on the deflector plane and the geometry of the
  source-deflector-observer optical system are established.  This is
  done by numerically simulating the lensing of light past a parametric
  mass model, and adjusting the handful of model parameters to match the
  positions and shapes of the observed and simulated lensed arcs.  At
  the same time, this determines the magnification of the background
  source induced by the lensing process.  The predicted magnification is
  then removed from the data to reveal the intrinsic, though still
  distorted, background distribution of light.  After tracing each
  lensed ray back to the source plane, the data are combined to produce
  a surface brightness distribution of the source.  This two-stage
  inversion scheme produces a parametric model of the deflector and a
  pixelized rendering of the background source which together mimic the
  observed gravitationally lensed features.  We test the viability of
  scheme itself on a well-studied collection of lensed objects in the
  galaxy-cluster MS~2137.  Confident in the algorithm, we apply it
  second time to predict the distribution of mass in the galaxy-cluster
  MS~1455 responsible for an observed triplet of lensed arcs.  Our
  predictions about the lens in MS~1455 make it particularly
  interesting, for a single background source is responsible for both
  tangential arcs and a radial arc.
\end{abstract}



\section{Introduction}

The study of gravitational lensing has evolved from a novel application
of General Relativity to an astronomical tool, for the analysis of
lensing provides an additional, independent measure of the mass
distribution of galaxies and clusters of galaxies.  Comparison of the
lensing mass and luminous mass, for instance, can begin to answer
questions about the nature of the dark matter in these objects.

One way to extract the information encoded in the strong lensing
behaviour of a cluster of galaxies is to produce a model of the mass
distribution which, together with one or more luminous, background
sources, produces the observed collection of lensed arcs and arclets.
Several lens inversion schemes have been developed in the last 10 years
to study lenses characterized by sets of compact lensed arcs, each set
being multiple images of a background source.  The complexity of the
schemes and the strength of their predictions have increased along
with advances in observations.  

Kochanek \& Narayan (1992) developed the Lens\-Clean algorithm, based
on Kochanek's earlier Ring Cycle (1989), to study lensing systems
containing extended images, particularly Einstein Rings formed from
background radio sources.  The result is a discrete map of the mass
distribution on the lens plane.  The LensMEM routine of Wallington et
al. (1996) introduces the maximum entropy method (MEM) into the
inversion routine.  Parametric model parameters are adjusted so that the
background source needed to reproduce the observed lensed features is
the ``most probable...consistent with the data''\footnote{Narayan \&
  Nityananda 1986, 128.} and therefore the most
``natural.''\footnote{Ibid., 137.}

Another family of lens inversions schemes is based upon parametric lens
models where the nature of the deflector is specified, and only the
values of the parameters are altered.  These methods are based on the
fact that when multiple images of a common background source are traced
back through the parametric lens model, the pre-images must coincide on
the source plane.

Mellier, Fort, \& Kneib (1993) (hereafter M93) produce
a parametric model of the mass distribution of the core of the
galaxy-cluster MS~2137.  A large pseudo-isothermal, elliptical mass
distribution is postulated, and three lensed images are traced back to
the source plane.  Parameter values are selected by minimizing a
$\chi^2$ statistic measuring the distances between the three pre-images
on the source plane.  Lensing in the galaxy-cluster A2218 is examined,
first with ground-based data (\cite{KN95.1}) and later with \emph{HST}
observations (\cite{KN96.1}.)  In the latter, parameters describing four
large galaxies and 30 smaller galaxies are fit through seven
multiply-imaged background sources.  Nair (1998) generates a parametric
model reproducing the 10 lensed images observed in B$1993+503$ by
minimizing distances between pre-images on the source plane, while at
the same time demanding the lensed images show the correct parity.
Tyson, Kochanski, \& Dell'Antonio (1998) constrain a 512 parameter model
of CL~$0024+1654$ by matching close to 4000 lensed pixels in \emph{HST}
and numerically simulated images.  The stunning detail is the result of
a very expensive computation.

High-resolution \emph{HST} images of MS~2137 allow Hammer et al. (1997)
(hereafter H97) to produce a more complex parametric
model of the mass distribution of the cluster core, as well as
reconstructions of the background sources.  Values for the model
parameters are estimated by selecting bright knots of light observed
within multiple lensed images, a triplet of images of one source and a
pair of images of a second source, and adjusting the parameters to make
the pre-images of these knots coincide on the source plane.  Then the
lensed images in their entirety are traced back to the source plane to
reconstruct the two sources.

It is critical in the lens inversion schemes of Mellier, Fort, \&
Kneib, Kneib et al., Hammer et al., and Nair that common structures or
knots of light be identified in two or more lensed arcs.  The model
parameters are chosen by forcing these structures back to a common
origin on source plane.  Because of the extreme magnification that
occurs near the critical lines of the lens, faint regions of the
background which lie near the corresponding caustics may be greatly
magnified and appear as bright knots of light.  These knots can be
misidentified as coming from bright structures in the background source.
Adjusting the model parameters to make coincident these faint regions
and regions of the source which do show structure may lead to
inconsistent models of the lens.  This problem stems from the fact that
the appearance of a lensed arc is the product of both the structure of
background source and the effects of magnification of the lens.

The two-stage inversion algorithm is described in detail in Section~2,
using the lensing observed in MS~2137 as a test case. In this Section,
we introduce polar moments, statistics used to quantify the position and
shapes of lensed arcs.  In Section~3, the algorithm is applied to the
galaxy-cluster MS~1455, producing a model which suggests that a single
background source is responsible for both a tangential and a radial arc.
Our model does not fully reproduce the lensing behaviour of MS~1455,
revealing both shortcomings of the model and also the difficulties
associated with inverse problems.  In Section~4, we discuss strengths
and weakness of the new inversion scheme, particularly the use of polar
moments.   Finally, our conclusions are summarized in Section~5.

\section{A Two-Stage Inversion Algorithm}

We introduce a new inversion algorithm for producing a model of the
geometry of the gravitational lens, a parametric description of the
cluster mass distribution, and a reconstruction of the background
source.  The key to this new approach is decoupling the effects
of lens magnification and background source structure in the appearance
of multiple lensed arcs.  The algorithm is stated briefly here, and then
illustrated with the well-studied gravitational lens MS~2137.

The first stage of the inversion is to build a parametric model of the
deflector mass distribution, establish the redshifts of the lens and
source planes, and determine the position of the background source,
based only on the positions and shapes of the lensed arcs in the
observations.  As discussed below, the positions and shapes are
characterized by polar moments.  The model establishes a family of
``conduits'' through which the solutions to the lens equation pass.  At
the same time, the models fixes the magnification throughout the lens,
so that the magnification factor at each point on the deflector plane
can be calculated and removed from the data.

The second stage of the inversion algorithm is the reconstruction of the
background source responsible for the observed arcs.  Each pixel in the
observations identified as containing lensed light is traced back to the
source plane along the solution to the lens equation, and the
magnification factor is removed.  This produces a collection of points
on the source plane, each point carrying the flux of the
source as it would appear in the absence of lensing.  There are many
choices for interpolating this data across the source plane to build a
pixelized image of the source.  We adopt a simple strategy by choosing a
uniform pixel size, and then setting the pixel size so
that the number of pixels in the reconstructed source is comparable to
the number of data, namely the number of lensed pixels identified in the
observations.

As a test of consistency of the model, the reconstructed source is
passed back through the model to check for spurious structure within the
arcs, or spurious arcs altogether.

\subsection{Arcs in MS~2137}

To test the validity of the new lens inversion algorithm, the algorithm
is applied to the lensed objects observed in the galaxy-cluster MS~2137,
for which models have already been produced in M93 and H97.

\begin{figure}[ph]
  \plotone{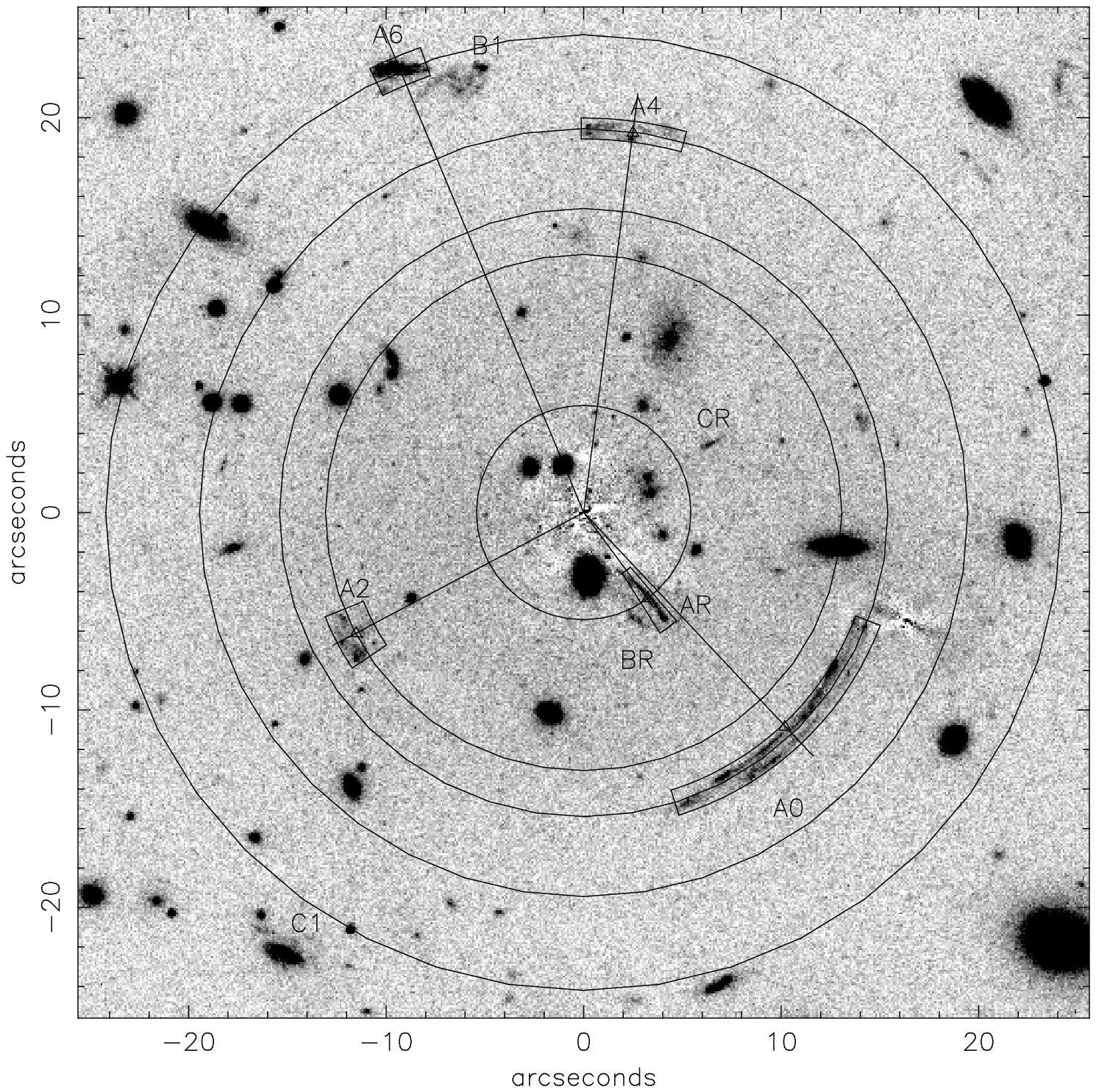}
  \caption{WFPC2 image of MS~2137, showing lensed arcs A0-A2-A4, AR-A6,
    BR-B1, and CR-C1.  The central cD galaxy and a smaller galaxy
    obscuring the upper end of arc A0 have been removed.  A circle of
    radius $\bar{r}$ and a line at position angle $\bar{\theta}$ are
    drawn through the polar centroid of each arc.  The shapes of the
    arcs are characterized by the annular sectors drawn around each
    arc. \label{fig:2137obs}}
\end{figure}

A \emph{HST} image of the centre of MS~2137 is shown in
Figure~\ref{fig:2137obs}.\footnote{Based on observations with the
  NASA/ESA Hubble Space Telescope, obtained from the data Archive at the
  Space Telescope Science Institute, which is operated by the
  Association of Universities for Research in Astronomy, Inc. under NASA
  contract No. NAS5-26555.}  The central cD galaxy and a smaller cluster
galaxy, identified as G1 and G7, respectively, in M93, have been removed
to reveal more structure of the arcs.  In previous models of MS~2137 and
in the model below, the giant tangential arc A0 and two counter-images
A2 and A4 are traced to one source in the background.  The radial arc AR
and its counter-image A6 are traced to a second source.  As noted in
H97, a third source produces the BR-B1 pair of arcs.  Our model also
predicts a fourth background source responsible for a further C1-CR pair
of arcs.

\placefigure{fig:2137obs}

To select values for the parameters in a parametric model of the lens,
the positions and shapes of numerically simulated lensed arcs are
``matched'' to the positions and shapes of the observed arcs.
Quantifying positions and shapes of lensed objects is the basis of the
inversion-via-distortion techniques applied to weak lensing inversions
(\cite{KA93.1}).  In weak lensing analyses, the (flux-weighted) centroid
and quadrupole moments of a weakly distorted background galaxy are
calculated, and the galaxy is modeled as an equivalent ellipse.  This
analysis cannot be applied directly in the strong lensing regime because
the arcs are not generally elliptical is shape (for example, the giant
arc A0.)  Instead, we introduce ``polar moments'' of the lensed images:
The polar coordinates $(\theta,r)$ of pixels containing lensed light are
interpreted as if they are coordinates in a Cartesian coordinate system.
By summing over pixels about a chosen threshold, $I_o$, the following
statistics are tabulated:

\begin{eqnarray}
  Q_o & = & N\ (\textrm{number~of~lensed~pixels}) \label{eqn:polarQo} \\[2ex]
  \bar{r} & = &
  Q_o^{-\!1}\sum_{I_i>I_o} r_i \label{eqn:polarrbar} \\
  \bar{\theta} & = & 
  Q_o^{-\!1}\sum_{I_i>I_o} \theta_i \label{eqn:polartbar}\\[2ex]
  Q_{rr} & = &
  Q_o^{-\!1}\sum_{I_i>I_o} (r_i-\bar{r})^2 \label{eqn:polarQrr} \\
  Q_{r\theta} & = &
  Q_o^{-\!1}\sum_{I_i>I_o} (r_i-\bar{r})(\theta_i-\bar{\theta}) 
  \label{eqn:polarQrt} \\
  Q_{\theta\theta} & = &
  Q_o^{-\!1}\sum_{I_i>I_o} (\theta_i-\bar{\theta})^2 \label{eqn:polarQtt}
\end{eqnarray}

\noindent
These moments are not flux-weighted, but depend only on the positions of
the lensed features on the image plane.  The $0$th moment $Q_o$ is the
area of the image, in units of $\Omega=\Delta^2$ arcsec$^2$, where
$\Delta$ is the arcsecond pixel length of the pixels in the observations.
The radial moment $\bar{r}$ specifies the average radius of the lensed
image, for there is just as much weight in the image outside the circle
of radius $\bar{r}$ as there is inside this circle.  In the case of
giant arcs, $\bar{r}$ should closely approximate the Einstein Radius of
the lens.  The moment $\bar{\theta}$ specifies the average position
angle of the lensed image: there is just as much weight clockwise from
the line at position angle $\bar{\theta}$ as there is counter-clockwise
from this line.  As $\theta$ simply measures position angle on the sky,
the magnitudes of $\bar{r}$ and $\bar{\theta}$ are incomparable.

The second polar moments of an arc characterize its shape: $Q_{rr}$
measures the radial spread of the arc, while $Q_{\theta\theta}$ measures
the angular spread.  Analogous to the equivalent ellipse of weak lensing
analyses, we construct a representative region based on the values of
these components.  A uniform rod of length $2L$ lying along the $x$-axis
between $-L$ and $L$ has a second moment $Q_{xx}=\frac{1}{3}L^2$.  The
length can be recovered from the moment, $L=\sqrt{3Q_{xx}}$.  We apply
this result to the polar moments of the lensed arcs.  The region lying
between \mbox{$\bar{r}\pm\sqrt{3Q_{rr}}$} and
\mbox{$\bar{\theta}\pm\sqrt{3Q_{\theta\theta}}$} is a rectangle in the
Cartesian system, and what we refer to as an ``annular sector'' in the
polar coordinate system.

The polar moments of the A0-A2-A4 and AR-A6 arcs are listed in
Table~\ref{tbl:2137datamoments}.  The quantities $r_i$ and $\theta_i$
which enter the polar moments in
Equations~\hbox{(\ref{eqn:polarrbar})-(\ref{eqn:polarQtt})} are simply
the coordinates of the centre of each pixel containing lensed light, and
are not based on the surrounding light distribution.  The radial
coordinate of any ray of light which strikes a lensed pixel is therefore
accurate only to $\delta r=\Delta/2$, which amounts to $0\farcs050$ for
the $0\farcs100$ resolution of the \emph{HST} observations.  An
uncertainty of $\Delta/2$ arcseconds at a radius of $\bar{r}$
corresponds to an uncertainty in position angle
\[
\delta\theta=
\frac{90\Delta}{\pi \bar{r}}\ \textrm{degrees}\ \ .
\]
The annular sectors built from the
quadrupole moments are shown in Figure~\ref{fig:2137obs}, where each 
annular sector sits at the intersection of a circle of radius $\bar{r}$
and a radial line at position angle $\bar{\theta}$.

\placetable{tbl:2137datamoments}

In weak lensing analyses, the coordinate frame in which the matrix of
quadrupole moments is diagonal defines the principal axes of the
equivalent ellipse.  In the polar moments scheme, the off-diagonal
moment $Q_{r\theta}$ cannot be interpreted so easily.  This
is due in part to the incomparable dimensions of the quadrupole moments.
We can extract some information, nevertheless, from the sign of the
$Q_{r\theta}$ moment.  Spherically symmetric arcs have equal
weight inside and outside the circle of radius $\bar{r}$, and equal
weight clockwise and counter-clockwise from the radial line at position
angle $\bar{\theta}$.  The off-diagonal moment $Q_{r\theta}$
in Equation~(\ref{eqn:polarQrt}) vanishes.  Outside (inside) the the
centroid circle, $r-\bar{r}$ is positive (negative); counter-clockwise
(clockwise) from the centroid radial line, $\theta-\bar{\theta}$ is
positive (negative).  Thus the sign of $Q_{r\theta}$ shows
any asymmetry of the image inside the annular sector.  A lensed image
rotated clockwise about the $(\bar{\theta},\bar{r})$ centroid, like arc
A6 in Figure~\ref{fig:2137obs}, more heavily populates the regions where
$Q_{r\theta}>0$.  Similarly, $Q_{r\theta}<0$ for
images rotated counter-clockwise with respect to the polar centroid,
like the giant arc A0.

Comparing the polar moments tabulated for the different types of lensed
arcs is revealing.  In the weak lensing regime, the ratio of the major
and minor axes of the equivalent ellipse gives a measure of the
ellipticity of the distorted background galaxy.  In the strong lensing
regime, we define a shape parameter $\chi$ by measuring the ratio of the
dimensions of the annular sector built from the polar quadrupole moments:
\[
\chi=
\frac{\textrm{tangential~dimension}}{\textrm{radial~dimension}} =
\frac{\pi\bar{r}\sqrt{%
    Q_{\theta\theta}}}{180\sqrt{Q_{rr}}}\ \ .
\]
A lensed feature which is circular produces $\chi\sim 1$.  In the case
of MS~2137, the giant tangential arc A0 shows $\chi\gg 1$, while the
radial arc AR shows $\chi<1$.  The other arcs in the collection are
tangentially distorted, with $\chi>1$ in each case.  The quantity $\chi$
may serve to distinguish between radial and tangential arcs, based only
on their polar moments.

The small collection of polar moments defined in
Equations~\hbox{(\ref{eqn:polarQo})-(\ref{eqn:polarQtt})} characterizes
the position of the lensed arcs quite well.  Four constraints can be
extracted from each image: $\bar{r}$, $\bar{\theta}$, $Q_{rr}$, and
$Q_{\theta\theta}$, or equivalently, $\bar{r}$, $\bar{\theta}$, the
shape parameter $\chi$, and one dimension, $\sqrt{3Q_{rr}}$. Including
the off-diagonal moment $Q_{r\theta}$ as a constraint is dubious,
although matching its sign between the observations and simulations
provides an additional check of the model.

\subsection{Models of MS~2137}

The arcs in MS~2137 are numerically simulated by ray-tracing through a
parametric model.  We model the dark-matter halo of cluster core with a large
pseudo-isothermal, elliptical mass distribution (PID).  The radial
profile of the mass density is given by
\[
\rho_{PID}(r)=\frac{\sigma^2}{2\pi G r_c^2}\,
\frac{1}{1 + (r/r_c)^2}
\]
where $\sigma$ is related to the line-of-sight velocity dispersion of
the distribution, and $r_c$ is the core radius of the distribution.  A
second, smaller mass distribution models the central cD galaxy,
following a profile proposed by Miralda-Escud\'{e} (1995):
\[
\rho_{cD}(r)=\frac{\sigma^2}{2\pi G r_c^2}\,
\frac{1+r/r_c}{( 1+r^2/r_h^2)^2}\ \ .
\]
The parameter $\sigma$ sets the mass of the cD, while the two scale
parameters $r_c$ and $r_h$ control the shape.  Both density profiles are
adapted to elliptical distributions following the prescription of
Schramm (1990).

The orientation and eccentricity of the cD are set to match observed
values, where eccentricity is defined as $\sqrt{1-(b/a)^2\,}$ where $a$
and $b$ are the semi-major and semi-minor axes of the ellipse,
respectively.  The scale parameters of the cD are fit to the observed
light profile through an iterative parameter estimation scheme.  The
mass parameter $\sigma$, essentially the mass-to-light ratio, is a free
parameter.  The centre, orientation, and eccentricity of the PID are
allowed to vary slightly from the central cD.  The core radius $r_c$ of
the PID is also a free parameter.  The mass of the PID $\sigma$ is set
to reproduce the observed line-of-sight velocity dispersion, following
the description of Binney \& Tremaine (1987) based on the Jeans'
Equation.  The redshift of the deflector plane is set to the observed
value of $z_d=0.313$; the redshift $z_s$ of the source plane is
predicted by the model.

The free parameters are chosen by simulating the A0-A2-A4 triplet of
arcs originating from a common source, S1.  As only the positions of the
lensed arcs are important at this stage, the lensed appearance of a
uniform, elliptical background source is simulated and the polar moments
of the resulting arcs are tabulated.  The interactive simulation program
immediately updates the lensing behaviour as the parameters are
adjusted: coordinates in multiples of $\Delta$, mass $\sigma$ in steps
of 25~km/s, redshifts in steps of 0.025, and each source's semi-major
axis in steps of $0\farcs05$, orientation in steps of $5\arcdeg$, and
eccentricity in steps of $0.02$.  The position of the source is
determined primarily by simultaneously matching the centroids
$(\bar{\theta},\bar{r})$ of the three arcs.  Values for the size,
eccentricity, and orientation of the background ellipse are set
primarily by matching the quadrupole polar moments of the three arcs.

\begin{figure}[ph]
  \plotone{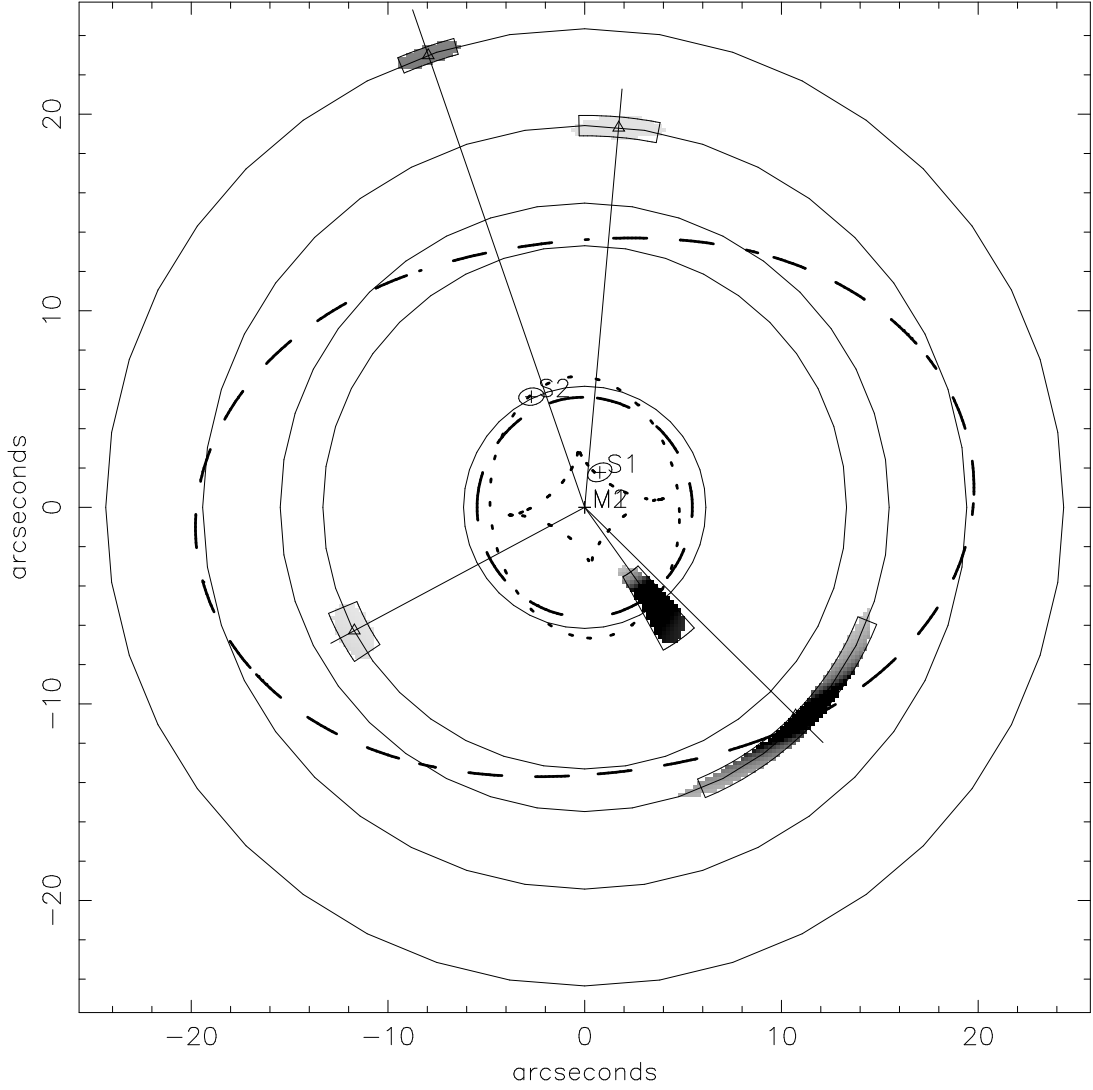}
  \caption{Simulation of MS~2137.  Dashed lines mark the critical lines
    of the lens; dotted lines trace the corresponding caustics.  The PID
    mass M1 and cD mass M2 coincide at the centre of the frame.  Small
    ellipses represent the positions of the two uniform, elliptical
    background sources S1 and S2.  For each of the five arcs, a circle
    of radius $\bar{r}$, a radial line at position angle $\bar{\theta}$,
    and an annular sector are drawn. \label{fig:2137pidcD2ELL}}
\end{figure}

The values chosen for the parameters are contained in
Table~\ref{tbl:2137paramtbl}.  The values are comparable to those found
in M93 and H97, also listed in the Table.  The simulation based on these
model parameters is shown in Figure~\ref{fig:2137pidcD2ELL}.  Annular
sectors around the simulated arcs are built from the moments listed in
Table~\ref{tbl:2137modelmoments}.  The polar moments of the simulated
arcs closely reproduce the observed moments of the A0-A2-A4 arcs.  In
particular, the shape parameter $\chi=11.4$ identifies arc A0 as a giant
tangential arc, $\chi=0.3$ identifies AR as a radial arc, and the sign
of the moment $Q_{r\theta}$ correctly characterizes the
asymmetry of each arc.
 
\placetable{tbl:2137paramtbl}

\placefigure{fig:2137pidcD2ELL}

\placetable{tbl:2137modelmoments}

To further justify the choice of parameter values, a second source S2 is
added at the same redshift as S1 (following H97) without changing the
deflector mass distribution, to test the ability of the model to
reproduce the AR-A6 pair of arclets.  The second source and its two
lensed images are included in Figure~\ref{fig:2137pidcD2ELL}.  The close
match between the polar moments of the observed and simulated AR-A6 arcs
in Tables~\ref{tbl:2137datamoments} and~\ref{tbl:2137modelmoments}
supports our selection of model parameters.

The two-mass, one-source model of MS~2137 is described by 20 parameters
listed in Table~\ref{tbl:2137paramtbl}.  The redshift of the deflector
plane, and the centre, orientation, eccentricity, and scale lengths of
the cD galaxy are deduced from observations, leaving 13 free parameters.
The three arcs A0, A2, and A4 provide 12 constraints on the model,
leaving a one parameter family of models.  Because the constraints do
not directly measure the model parameters, there is not a particular
parameter that can identified as the free parameter. Instead, some
mixture of parameters varies in the family of models, for instance the
product of distance and mass which enters the lens equation.  The
addition of a second background source S2, assumed to lie at the same
redshift as S1, requires only 5 more parameters, while producing 8 data
from the two new lensed features.  Therefore, more information can be
extracted from the model that is needed to produce it, and the results
become predictive.  The geometry of the model provides a ready
explanation for lensed objects in the simulation which are independent
of those used to constrain the model in the first place.

\subsection{Reconstruction of the Sources}

Uniform elliptical disks are used to model the background sources in the
first stage of the lens inversion.  With the lens geometry and deflector
mass distribution established, these idealized sources can be replaced
with distributions reconstructed from the data itself.  The model for
the lens specifies the origin on the source plane of each lensed pixel
in the observations.  Furthermore, the model parameters determine the
magnification of the background at any point of the lens.  To
reconstruct the appearance of the background source(s), each pixel
containing lensed light in the observations is traced back to the source
plane along the solution to the lens equation and the magnification
factor is removed from the data.  This leaves a point on the source
plane carrying the surface brightness of the source as it would appear
in the absence of lensing.  A coherent picture of the background source
is constructed by interpolating between these points.

Data are uniformly spaced in the observations and cover the image plane
exactly once.  Because of the lensing distortions, data on the source
plane do not inherit this simple structure, but cluster about the
caustics of the lens.  We choose a uniform pixel size, $\Delta_s$, to
reconstruct a pixelized rendering of the background source.  This is the
simplest choice, and can surely be improved by exploiting the
concentrations of data.

Several choices are available to set the size of the source pixels.
Source pixels representing the same physical size on the source plane as
the pixels in the observations represent on the deflector plane cannot
contiguously cover the background plane, for the background is
physically larger than the foreground.  By choosing source pixels with
the same angular size as the pixels in the observations, the background
plane may be covered, but these source pixels extrapolate the
information in the data over a much larger region.  We adopt a strategy
which is a compromise between these two choices:  The size of the source
pixels is set so that the total number of pixels in the reconstructed
source is comparable to the number of data.  The total number of pixels
is used, not just those containing a signal, because blank (dark)
regions on the source plane may be just as important as regions
containing light.

The source plane is divided into a uniform grid, and a flux is assigned
to each source pixel following these steps:
\begin{enumerate}
\item Source pixels which are not pierced by any backwards-traced rays
  are assigned a value of NaN, and appear in the result as pixels of
  zero flux.
\item A source pixel pierced by a single backwards-traced ray carrying
  an observed signal $I$ from a point where the magnification is $\mu$
  is assigned the de-magnified flux, $S=\mu^{-\!1}I$.
\item If $k$ backwards-traced rays, each carrying signal $I_i$, error
  $\sigma_i$, and magnification $\mu_i$, pierce the same source pixel,
  the pixel is assigned a flux $S$ which minimizes the error
  \[
  \phi=\sum_{i=1}^k
  \left( \frac{I_i - |\mu_i|S}{\sigma_i} \right)^2\ \ .
  \]
\end{enumerate}

\begin{figure}[ph]
  \plotone{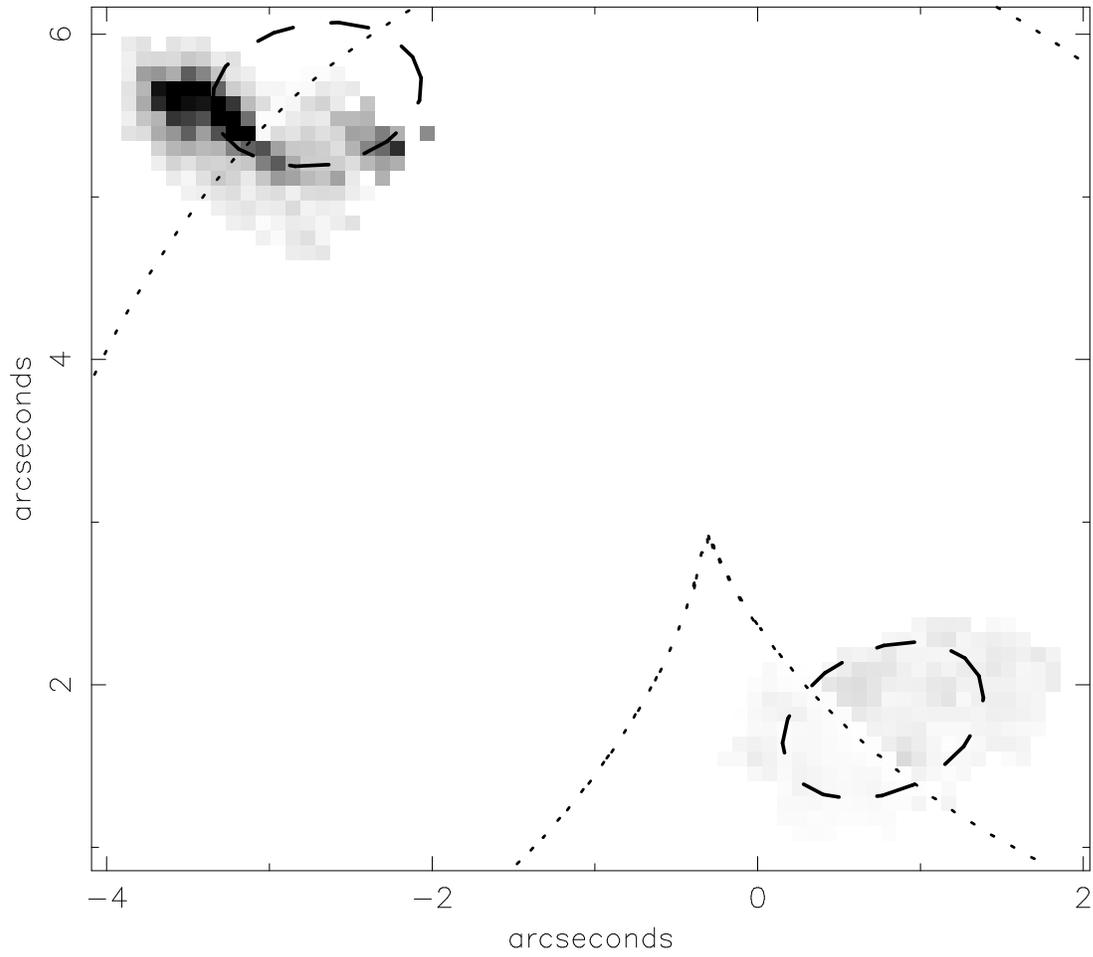}
  \caption{Reconstruction of the two sources behind MS~2137, S1 at the
    bottom right and S2 at the top left.  Dashed ellipses mark the
    location of the uniform elliptical disks used in the simulations.
    Both sources lie on caustics of the lens, traced with dotted lines.
    There are 3400 pixels in the frame, comparable to the 3350 data in
    the observations. \label{fig:2137S1andS2}}
\end{figure}

In the observations of MS~2137, we identify 2800 pixels in arcs A0-A2-A4
coming from source S1, and 550 pixels in arcs AR-A6 coming from source
S2, for a total of 3350 pixels containing lensed light.  The two sources
reconstructed from the 3350 pixels, following the strategy outlined
above, are shown in Figure~\ref{fig:2137S1andS2}.  The reconstruction of
source S1 closely coincides the position of the elliptical disk used to
simulate the arcs in Figure~\ref{fig:2137pidcD2ELL}.  The source lies
on the astroid-shaped tangential caustic, producing the giant tangential
arc A0.  The reconstruction of source S2 coincides with the second
elliptical disk added without altering the deflector mass parameters to
check the consistency of the model.  The second source crosses the
radial caustic, producing the radial arc AR.
Figure~\ref{fig:2137S1andS2} contains 3400 pixels with length
$\Delta_s=0\farcs092$, slightly smaller than the $0\farcs100$ 
resolution of the \emph{HST} data.  The majority of the pixels in the
reconstruction contain no signal, indicating the absence of additional
background light on this source plane which could form additional arcs
in the observations.  This reconstruction is comparable to that shown in
H97.

\placefigure{fig:2137S1andS2}

\begin{figure}[ph]
  \plottwo{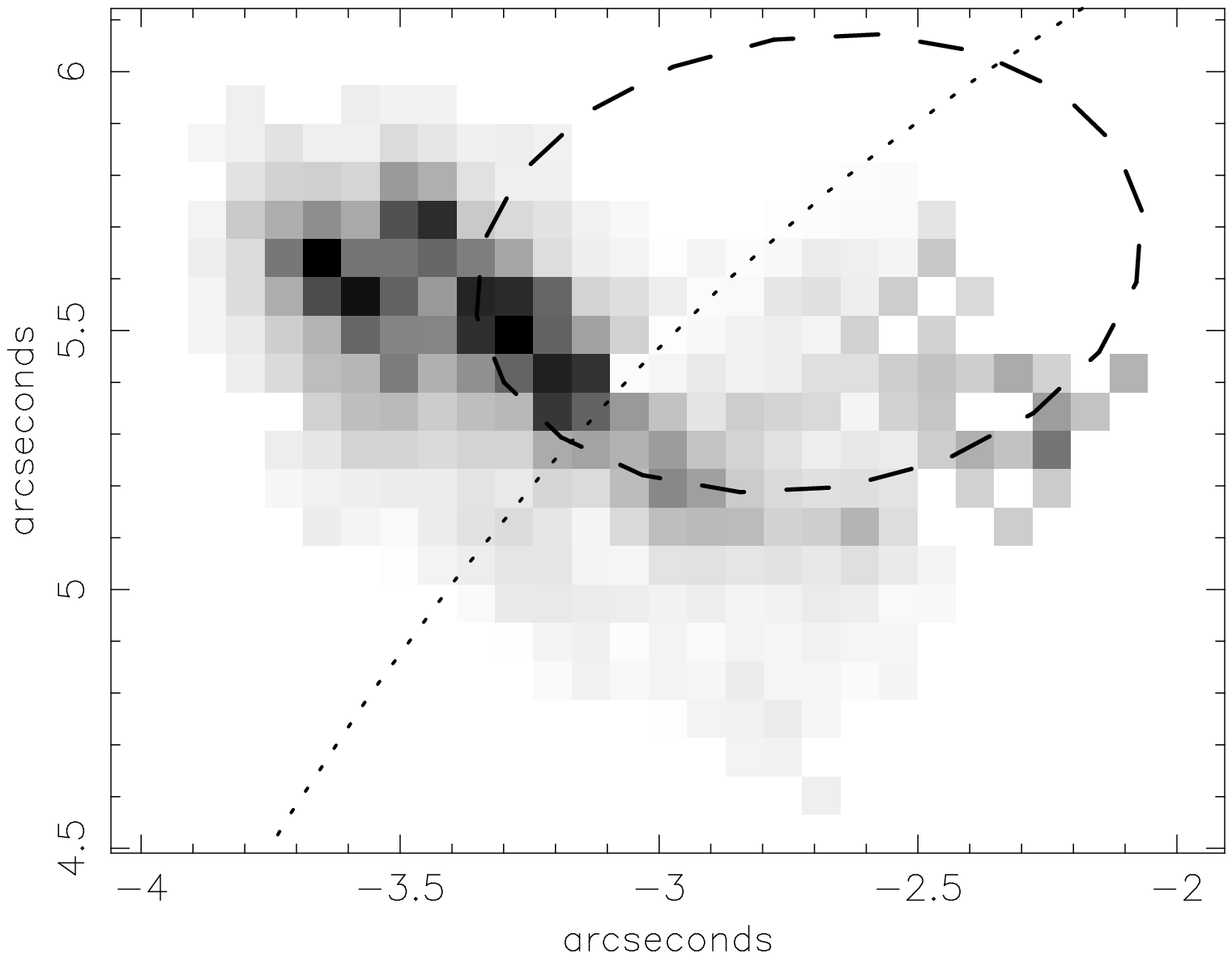}{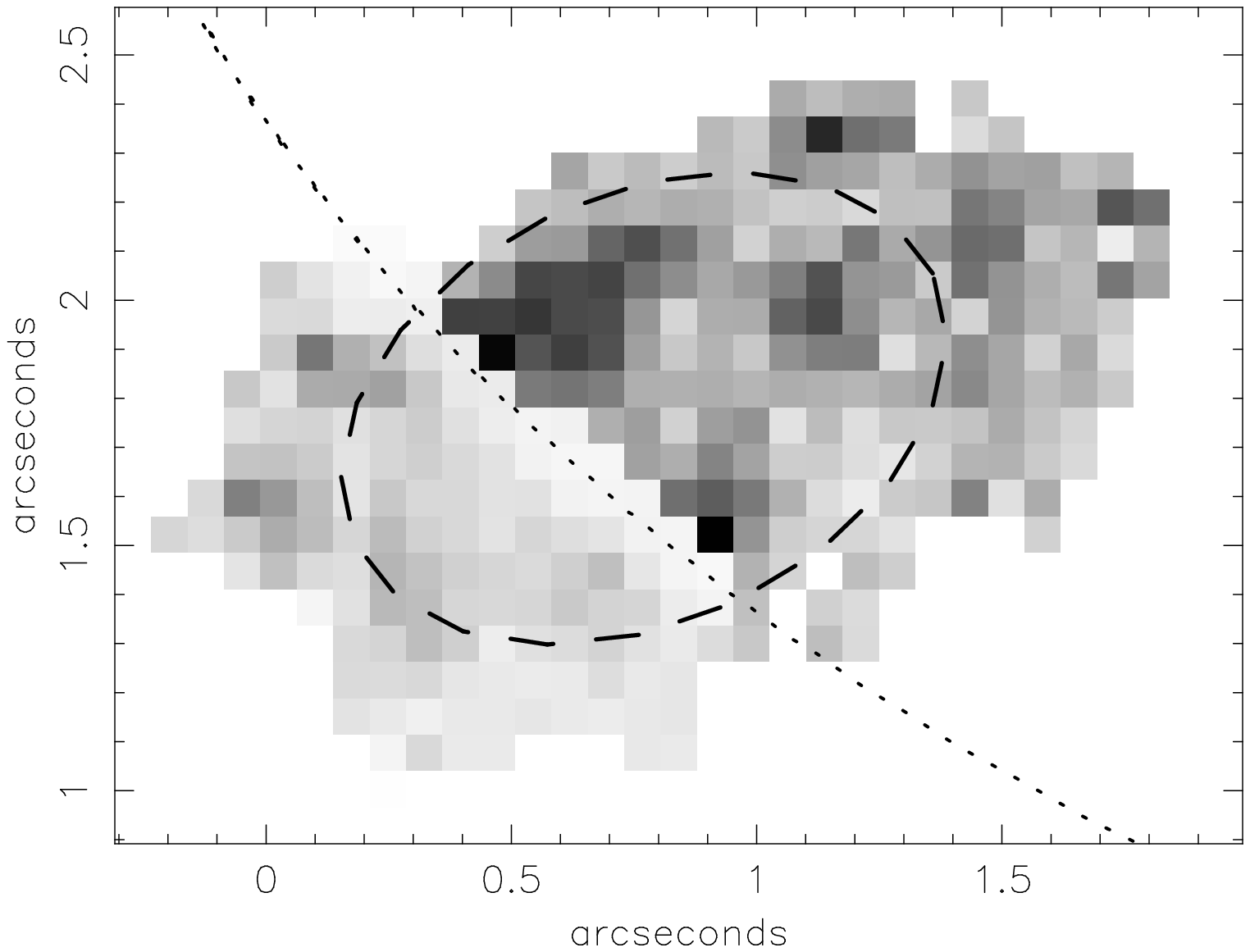}
  \caption{(left) Reconstruction of source S2.  The 550 data in arcs AR and
    A6 are interpolated onto 450 source pixels.  (right) Reconstruction
    of source S1.  The 2800 data in arcs A0, A2, and A4 are interpolated
    onto only 513 source pixels because of the low density of data
    coming from the weakly distorted A2, A4 arcs. \label{fig:2137S1orS2}}
\end{figure}

To better explore their internal structure, the two sources are
reconstructed separately in Figure~\ref{fig:2137S1orS2}.  In the
reconstruction of source S2, there are 450 pixels, with length
$\Delta_s=0\farcs074$,comparable to the 550 data drawn from arcs AR and
A6.  There are only 513 pixels with length $\Delta_s=0\farcs074$ in
the reconstruction of source S1, despite the 2800 data coming from the
A0-A2-A4 triplet of arcs.  The reason for this discrepancy is that arcs
A2 and A4 barely support a reconstruction at this resolution as they are
only weak distortions of the data sampled at $0\farcs100$.  The data
arriving from the giant arc A0 samples the source plane at a much higher
density because of the great distortion that occurs.

\placefigure{fig:2137S1orS2}

The reconstruction of source S1 in Figure~\ref{fig:2137S1orS2}~(right)
shows a curious faint stripe which follows the caustic of the lens.  It
is inconceivable that the source truly has a dim region so perfectly
aligned with the caustic, so the stripe must be a result of the modeling
process.  The numerical simulation of the lens in
Figure~\ref{fig:2137pidcD2ELL} shows a peak in the brightness of arc A0
where the image crosses the tangential critical line and the
magnification diverges.  The observations of arc A0 in
Figure~\ref{fig:2137obs} show the arc is very nearly uniform in
brightness all the way along its length, however.  The lack of a bright
peak in the data along the critical line results in a dim reconstruction
along the caustic.  The stripe is also a result of the finite resolution
of the simulation.  The magnification is infinite along the critical
line, and must be approximated.  We impose an upper limit of $100\times$
magnification in the reconstruction routine.  The approximation affects
any pixel in observations through which the critical line passes.  By
running the simulation at twice the resolution, the chain of effected
pixels remains, but with only one half the width.  The effect remains at
all finite resolutions, with the stripe becoming narrower and narrower.
The overall change in brightness between the parts of the source inside
and outside the caustic is due to the subtraction of galaxy G7 from the
data.

\subsection{Reconstruction of MS~2137}

\begin{figure}[ph]
  \plotone{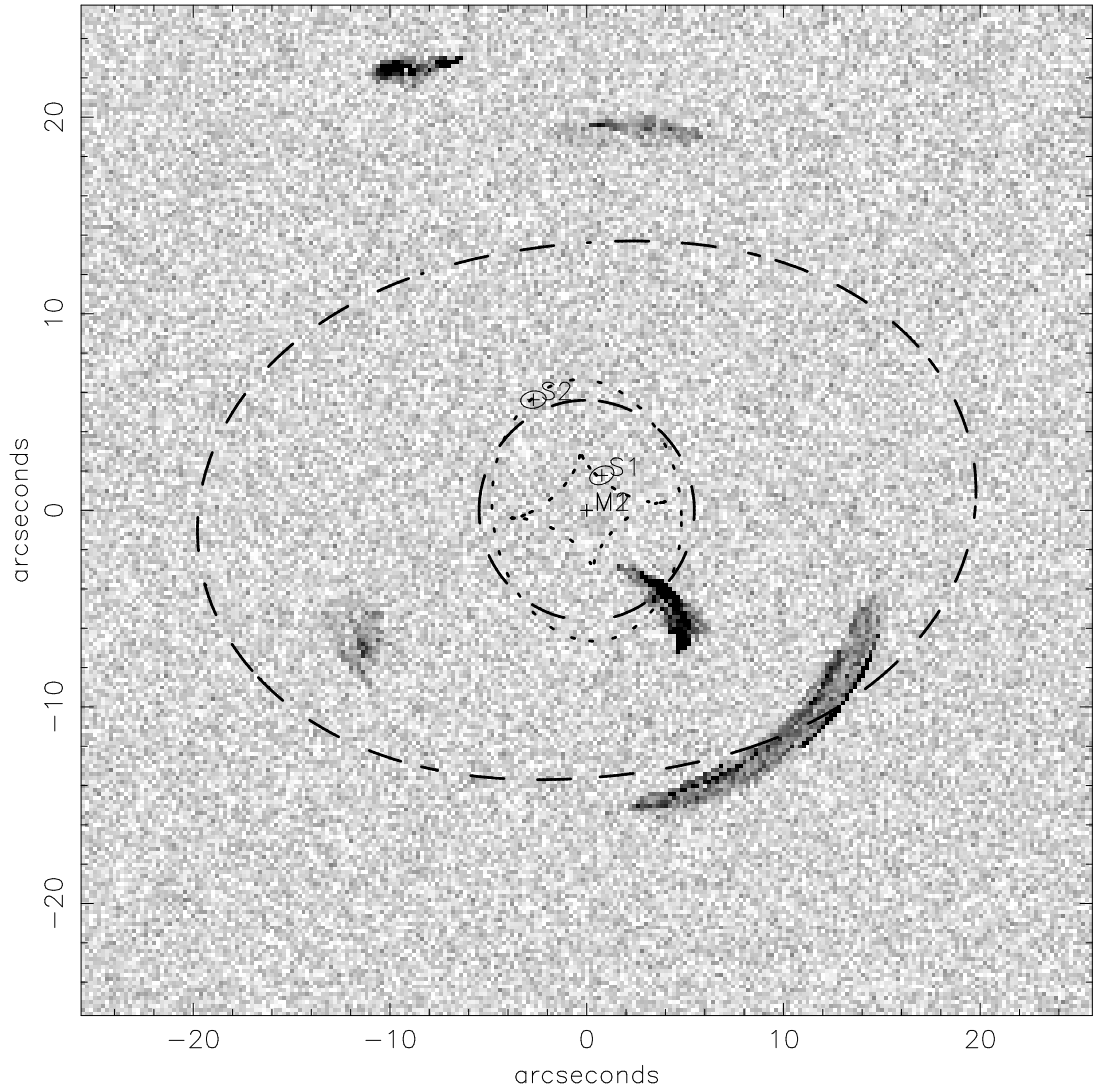}
  \caption{The reconstructed sources are passed back through the lens,
    and noise mimicking the \emph{HST} observations is added.  Five arcs
    are reproduced. \label{fig:2137relensed}}
\end{figure}

As a final test of the consistency of the model, the reconstructed
sources shown in Figure~\ref{fig:2137S1andS2} are passed back through
the parametric model.  The result is shown in
Figure~\ref{fig:2137relensed}, where gaussian noise matching that in the
data has been added. It is impossible to compare this Figure with the
observations at a pixel-by-pixel level without a complete model of the
sky and a thorough understanding of the noise.  It is apparent, though,
that the prediction is consistent with the observations.  We note in
particular (i) the reproduction of brighter knots in arcs A2, A4, and
A6, (ii) the twist in the radial arc AR, (iii) the double-ring structure
in the giant arc A0, and (iv) the absence of any extraneous lensed
objects.

\placefigure{fig:2137relensed}

In an ideal model, the re-lensed source perfectly reproduces the
observations.  Imperfections in the model are doubly amplified, though,
once in each direction through the lens.  The success of the results
provides compelling evidence that gravitational lensing, at the level
prescribed by this PID model, is actually occurring in the
galaxy-cluster MS~2137.

More importantly, the results show that the two-stage inversion
algorithm used to reconstruct MS~2137 is consistent with
other algorithms that exist today.  Fitting the lens with polar moments
to decouple the effects of magnification and source structure in the
observed arcs appears to be viable, at least in the cases of relatively
simple mass distributions with a well defined centre-of-lensing.  With
this confidence, we turn to the collection of features attributed to
gravitational lensing visible in the galaxy-cluster MS~1455.

\section{A Radial Arc in MS~1455}

\begin{figure}[ph]
  \plotone{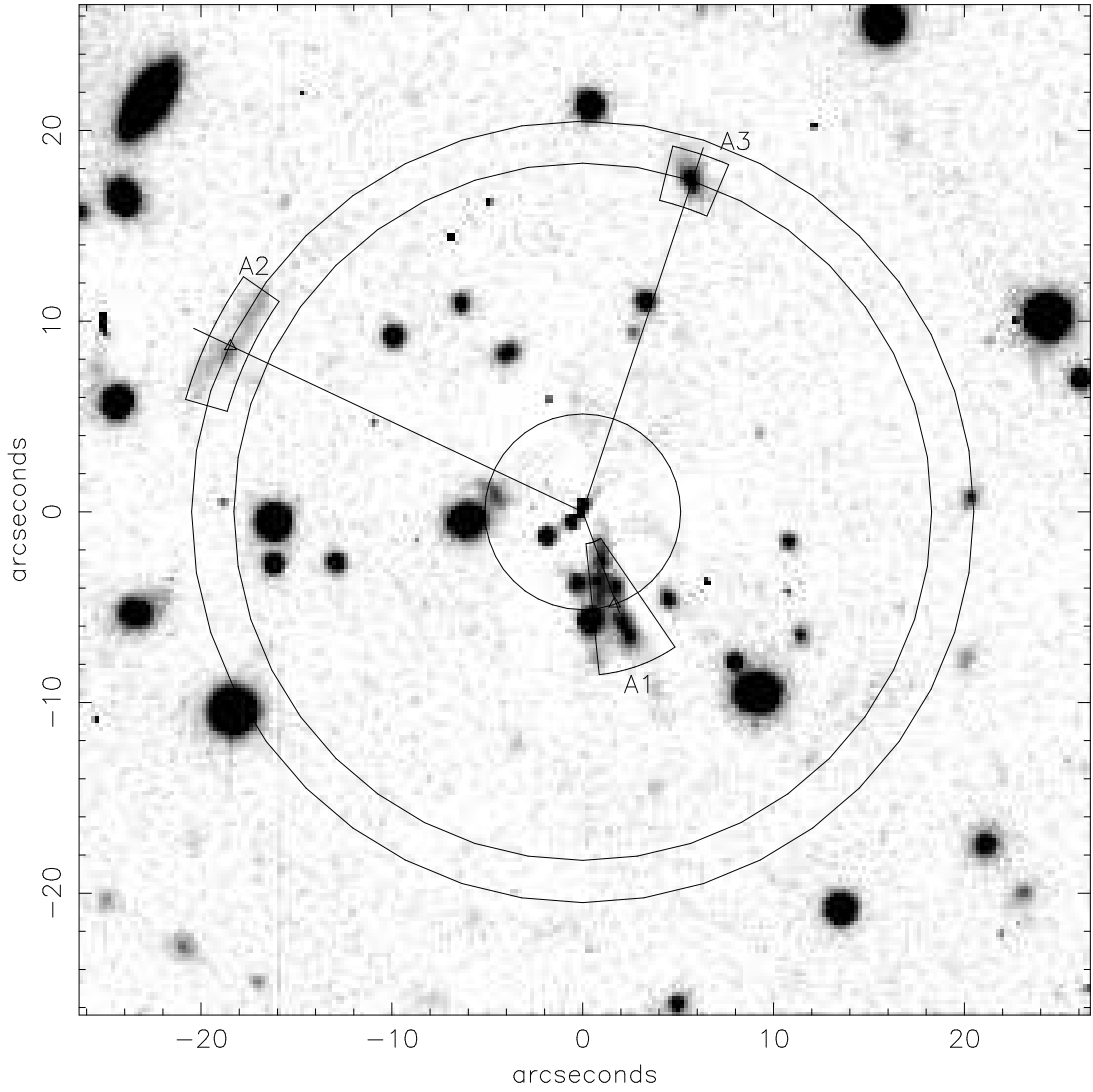}
  \caption{MS~1455 with the central cD galaxy removed, revealing a candidate
    radial arc A1.  Subsequent modeling of the lens suggests the object
    A3 is also a gravitationally lensed image.  Circles and annular
    sectors show the polar moments of the
    arcs. \label{fig:1455datamoments}} 
\end{figure}

The galaxy-cluster MS~1455+22, observed as part of the Einstein Medium
Survey of X-ray clusters, lies at redshift $z=0.257$.  A candidate
gravitationally lensed tangential arc was identified by LeF\`{e}vre et
al. (1994).  This prompted subsequent observations in
May, 1995 at the Canada-France-Hawaii Telescope (CFHT) as part of a weak
lensing survey of the cluster over a wide field of view.  The core of
the cluster appears in each of 12 overlapping 20-minute exposures.
These frames are aligned and added with IRAF routines, resulting in an
equivalent 4-hour exposure of the cluster core.  A hint of a structure
is visible in the envelope of the luminous central cD galaxy.  When the
cD galaxy is digitally removed, a collection of objects surrounding the
core is revealed, as shown in Figure~\ref{fig:1455datamoments}.  These
include several small cluster galaxies and an irregular radial feature
labeled A1 in Figure~\ref{fig:1455datamoments}, which we propose is a
radial arc.  The previously identified tangential arc is labeled A2.
During the initial modeling phase of this gravitational lens, a third
arc appeared in the simulations which closely matched the position of a
third diffuse object in the observations.  This arc, labelled A3 in
Figure~\ref{fig:1455datamoments}, is incorporated into the modeling
strategy.

\placefigure{fig:1455datamoments}

The lensed features in MS~1455 are similar to those seen in MS~2137.
Both clusters contain a radial arc and a large tangential arc.  The arcs
in MS~2137 appear in two sets, the A0-A2-A4 triplet due to source
S1, and the AR-A6 pair due to source S2.  Our analysis suggests
that the three arcs in MS~1455 are images of the same background source.
Radial and tangential arcs are produced across the two different types
of critical lines, which implies the single background source lies under
both the tangential and radial caustic.  As the caustics cross at only a
limited number of points on the source plane, this greatly constrains
the geometry of the lens.

From the position of 716 pixels in the observations containing lensed
light, polar moments are tabulated for the three arcs, listed in
Table~\ref{tbl:1455datamoments}.  Pixels in the CFHT data are
$\Delta=0\farcs207$ in length, producing an uncertainty of about
$0\farcs1$ in the radial positions.  The shape parameter $\chi$ again
distinguishes the radial arc A1 ($\chi=0.4<1$) from the tangential
arclet A2 ($\chi=3.1>1$).  The proximity of the radial arc to the
centre-of-lensing produces a wide angular width
$Q_{\theta\theta}$. The annular sectors built from the
moments are included in Figure~\ref{fig:1455datamoments}.

\placetable{tbl:1455datamoments}

\subsection{Parametric Models of MS~1455}

A simple model of MS~1455 consists of a large mass distribution to model
the halo of the cluster core, together with a smaller cD distribution at
the centre, and a single background source.  To begin to answer more
astrophysical questions, we build two models with two different halo
profiles.  The first model contains a PID mass, while the second model
uses a singular mass density profile proposed by Navarro, Frenk, \& White
(1995):
\[
\rho_{NFW}(r)=\frac{\sigma^2}{2\pi G r_s^2}\,
\frac{1}{(r/r_s)( 1 + r/r_s)^2}\ \ .
\]
The NFW density diverges as $r^{-\!1}$ at the origin and falls off as
$r^{-3}$ for $r\gg r_s$.  The profile has a well-founded basis in the
results of large numerical $N$-body simulations of cold dark matter.

In both the PID+cD and NFW+cD models, we allow the dark matter halo to
wander slightly from the cD galaxy in position, orientation, and
eccentricity.  The redshift of the lens plane is observed to be
$z_d=0.257$, while the redshift $z_s$ of the source is a free parameter.
The mass parameter $\sigma$ of the dark matter halo is set to reproduce
the observed line-of-sight velocity dispersion of the several dozen cluster
galaxies (\cite{CA96.1}).  The scale parameters of the cD are set to
match the profile of the surface brightness.  The values of the model
parameters we choose are listed in Table~\ref{tbl:1455paramtbl}.  The
significant difference in source redshift $z_s$ between the two models,
0.825 for the PID+cD model but only 0.620 for the NFW+cD model, may
serve to distinguish between the two if future observations are made.
Of the 20 parameters in the model, 8 are determined from the
observations, leaving 12 free parameters.  Three lensed arcs producing
12 constraints should be sufficient to constrain the parametric models
presented here.
  
\placetable{tbl:1455paramtbl}

\begin{figure}[ph]
  \plottwo{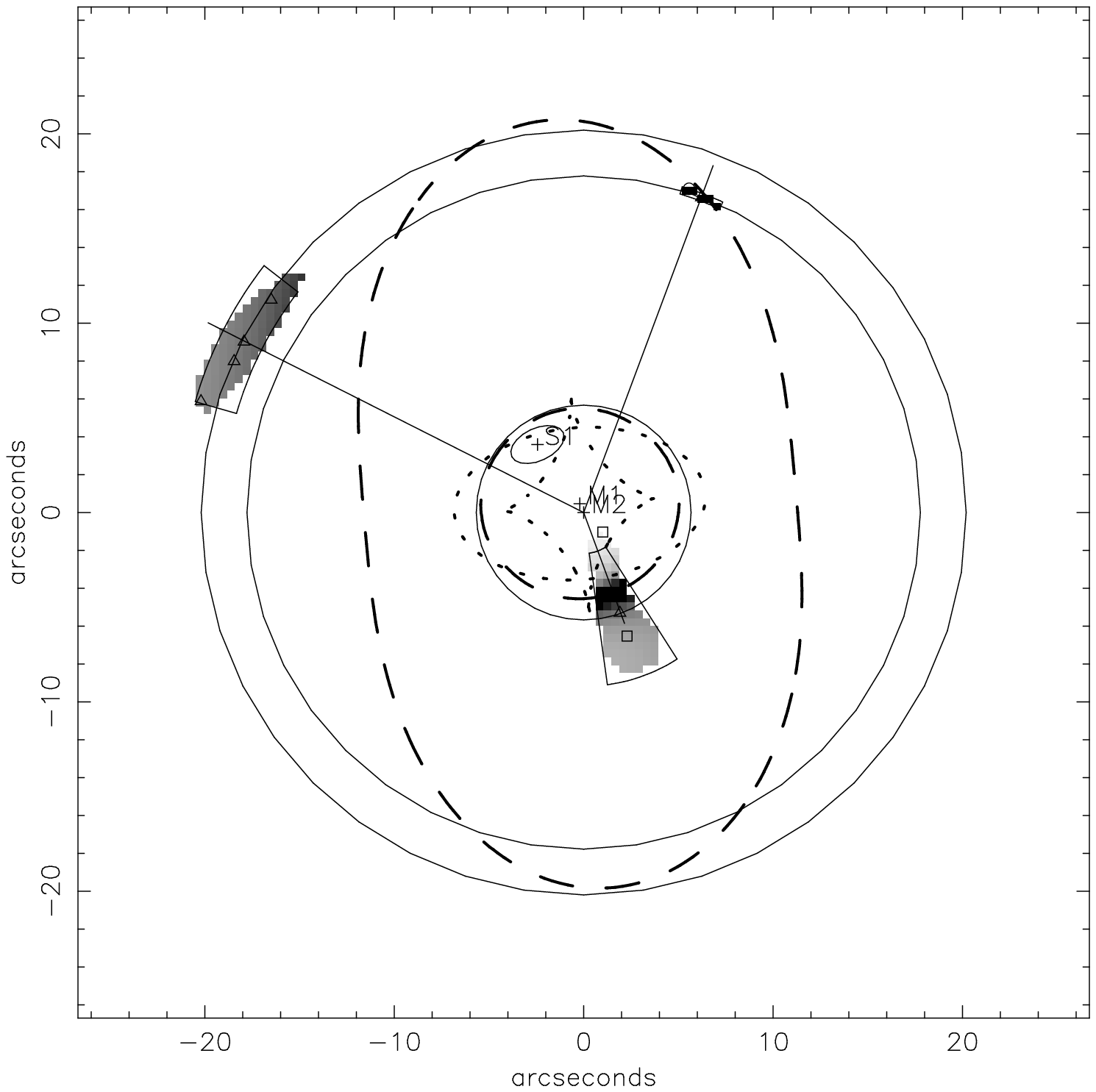}{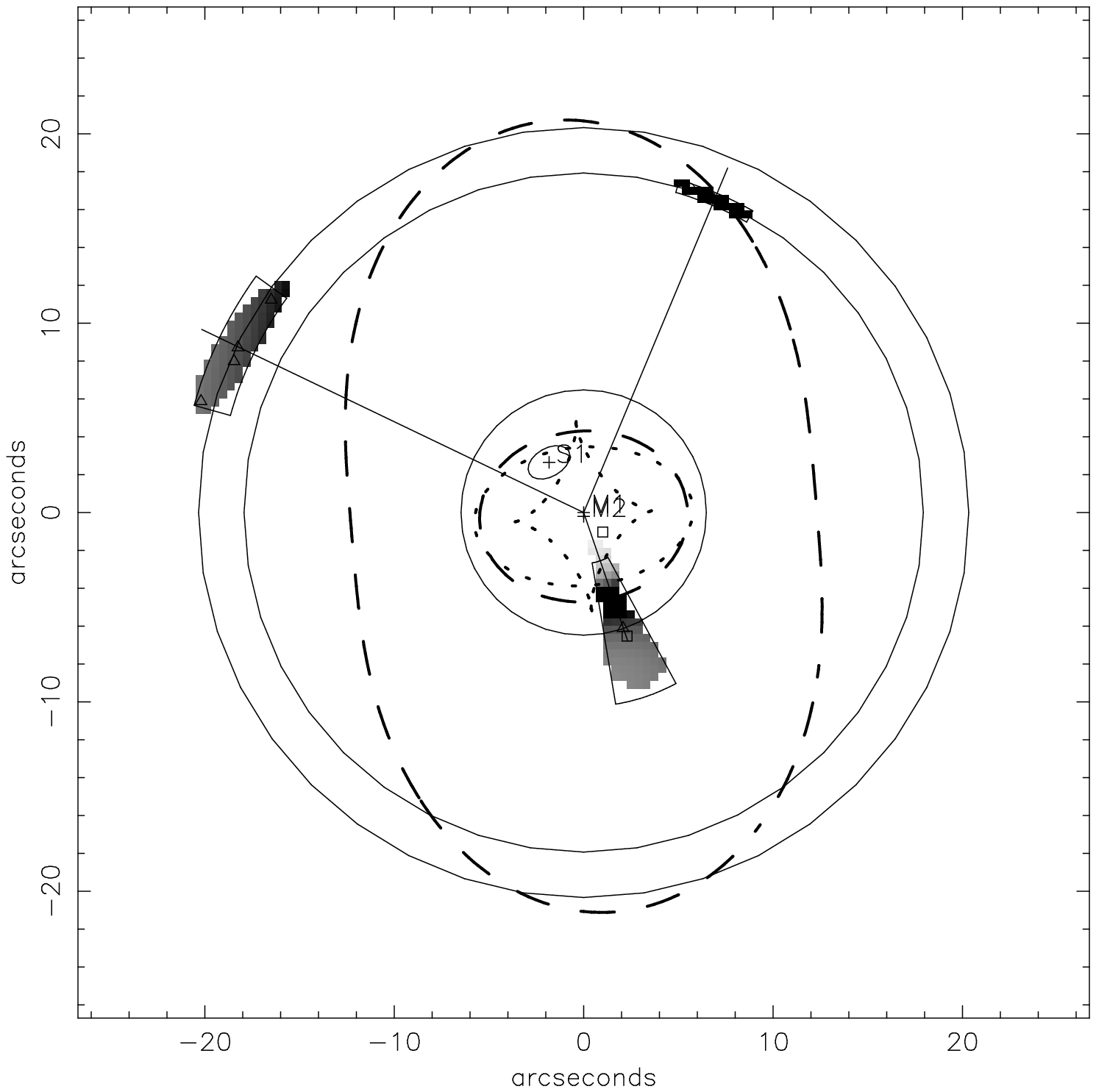}
  \caption{Simulations of MS~1455 with a single source behind a PID+cD
    lens (left) and the NFW+cD lens (right). \label{fig:1455pidandnfw}} 
\end{figure}

Figure~\ref{fig:1455pidandnfw} shows the numerical simulations of the
PID+cD and NFW+cD lenses.  Note how the single background source S1,
represented by a dashed ellipse, lies under both the astroid caustic
(forming arc A3) and the ovoid radial caustic (forming arc A1).  The
original tangential arclet A2 is the even parity image that forms
outside the network of critical lines.

\placefigure{fig:1455pidandnfw}

The polar moments of the arcs are listed in
Table~\ref{tbl:1455modelmoments}.  In both models of the lens, the
position of the tangential arclet A2 and the position angle of the
radial arc A1 are more carefully matched to the observations.  The polar
moments of the third image, arc A3, are treated more as a consistency
check.  Because of the extreme distortion that occurs near the critical
lines, the appearance of arc A3 in the simulations is very sensitive to
small changes in the parameters, particularly in the position and shape
of the background source.  The magnification across the critical lines
produces problems in the reconstruction discussed below.

\placetable{tbl:1455modelmoments}

\subsection{Source Reconstruction}

With the geometry of the lens fixed by the positions of the lensed arcs,
the magnification of the background source is determined.  Some 716
pixels in the observations are flagged as containing lensed light.  These
data are traced back to the source plane, and magnification factor is
removed.  

\begin{figure}[ph]
  \plottwo{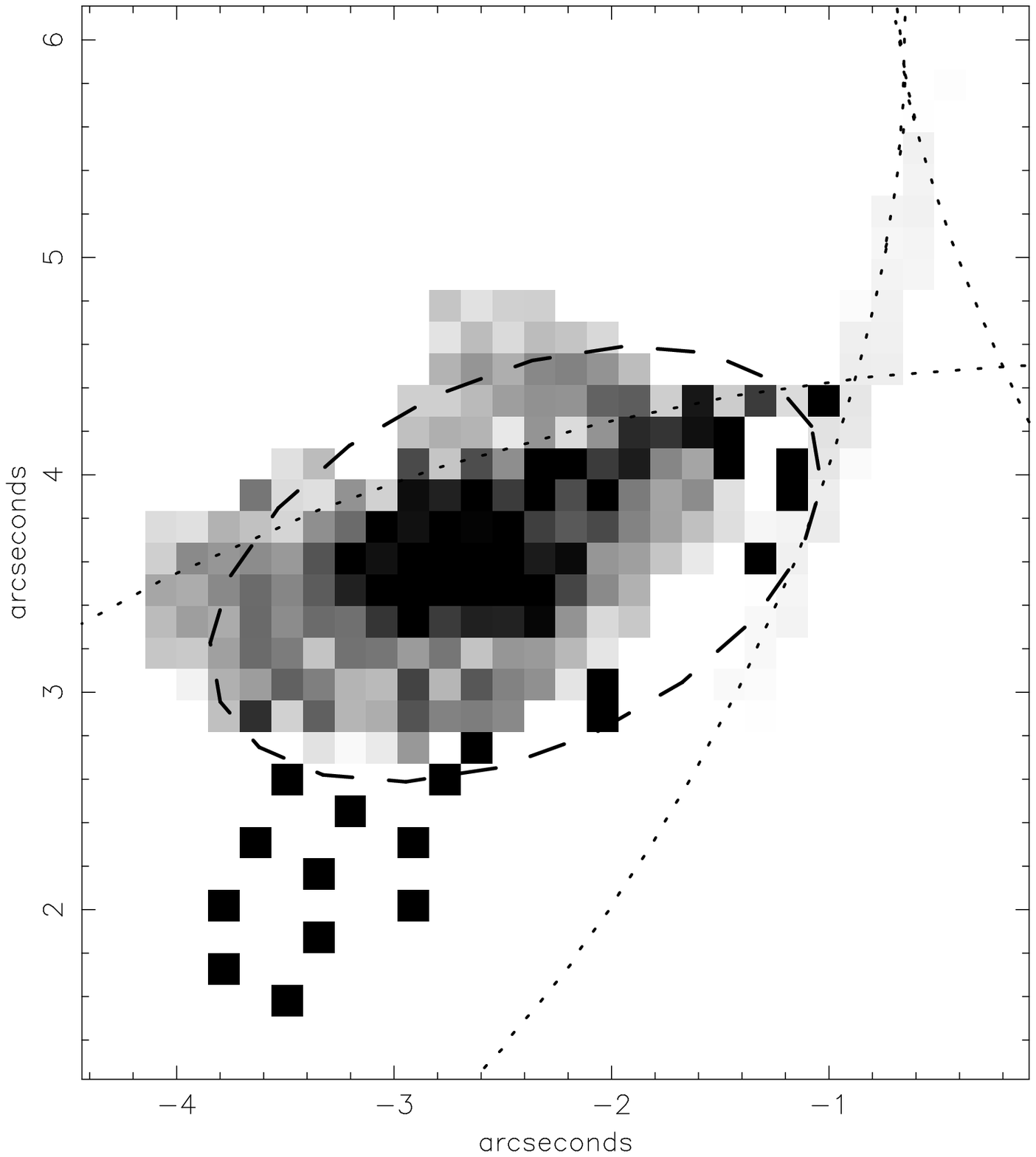}{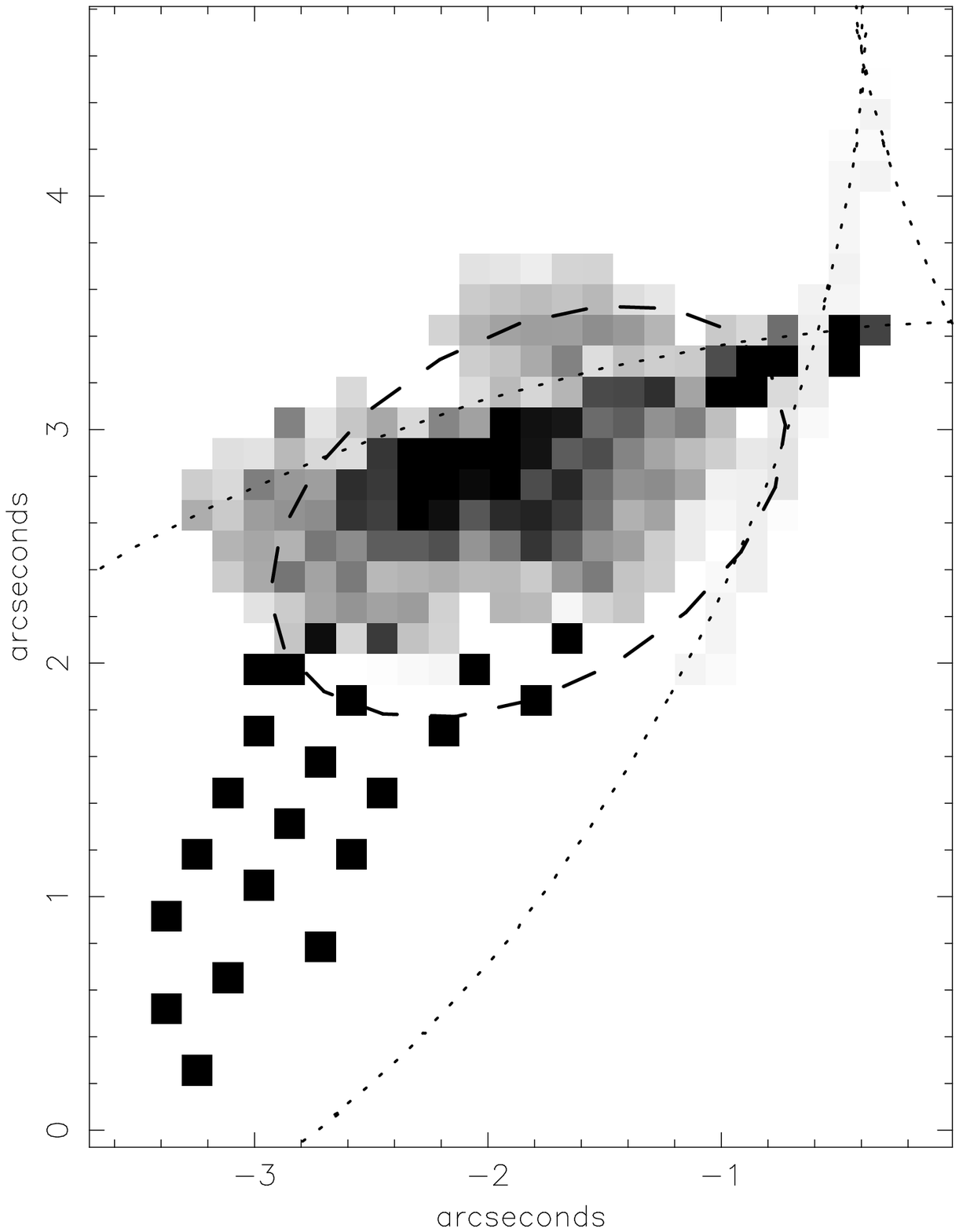}
  \caption{Source reconstructed behind the PID+cD lens (left) and
    the NFW+cD lens (right). \label{fig:1455S1}}
\end{figure}

The source reconstructed behind the PID+cD lens is shown in
Figure~\ref{fig:1455S1}~(left).  There are 780 pixels in the Figure with
length $\Delta_s=0\farcs145$, smaller than the $0\farcs207$ resolution
of the CHFT data.  Figure~\ref{fig:1455S1}~(right) shows the source
reconstructed behind the NFW+cD model from 792 pixels with length
$\Delta_s=0\farcs132$.

\placefigure{fig:1455S1}

Both reconstructions show the majority of the signal comes from a
generally elliptical object with a brighter central bulge, very likely a
spiral galaxy.  This galaxy closely coincides with the position of the
uniform elliptical disk used to simulate lensed arcs.  The data
contained in the third lensed arc A3 are wholly responsible for the
faint limb of the source which follows the astroid caustic to the
upper-right.  The flux is quite small because of the great magnification
the signal experiences as it passes through the lens.  It is likely the
source is surrounded by a low surface brightness extension, but only a
small portion of this can be seen through the high-magnification parts
of the lens.

\begin{figure}[ph]
  \plottwo{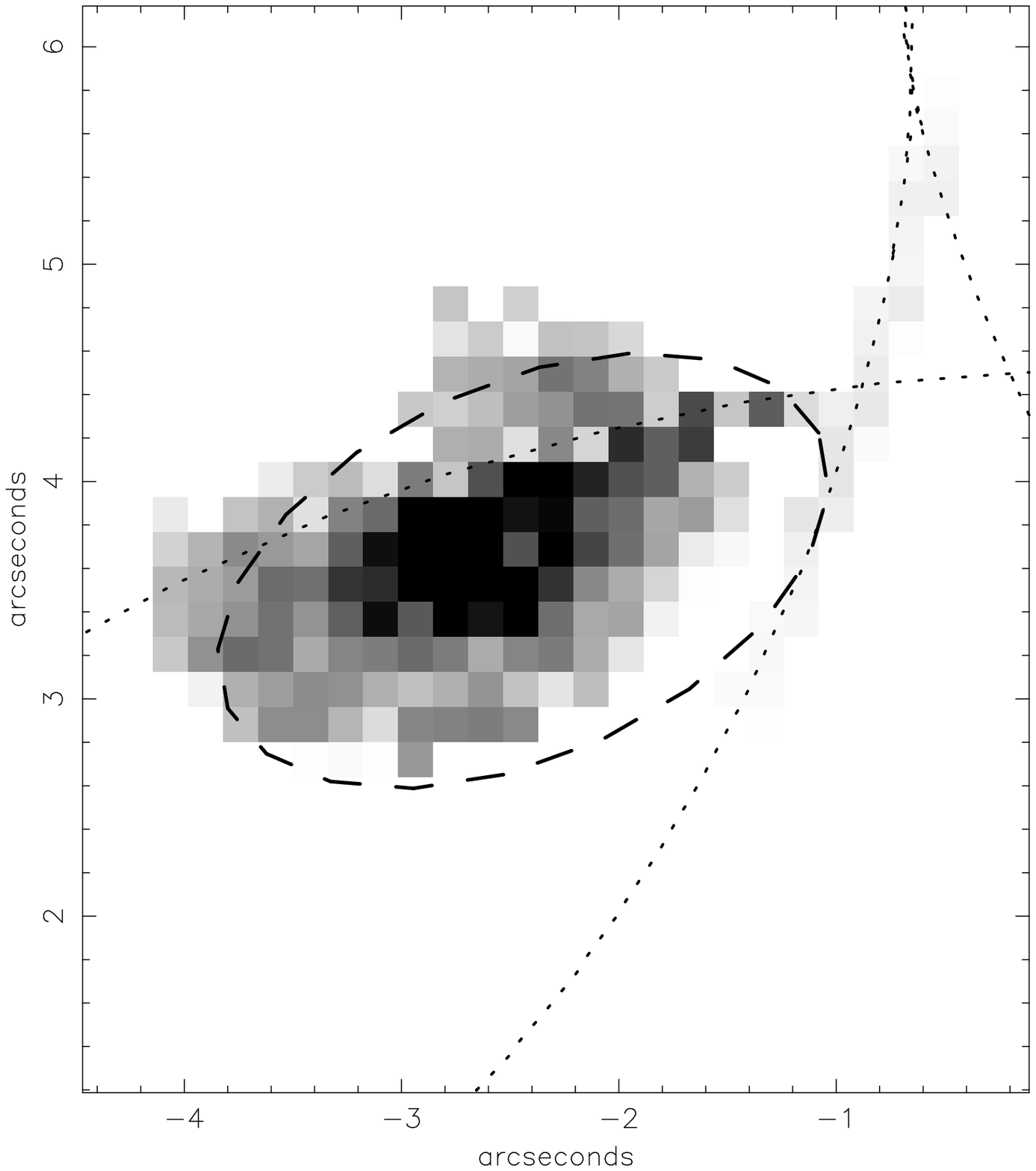}{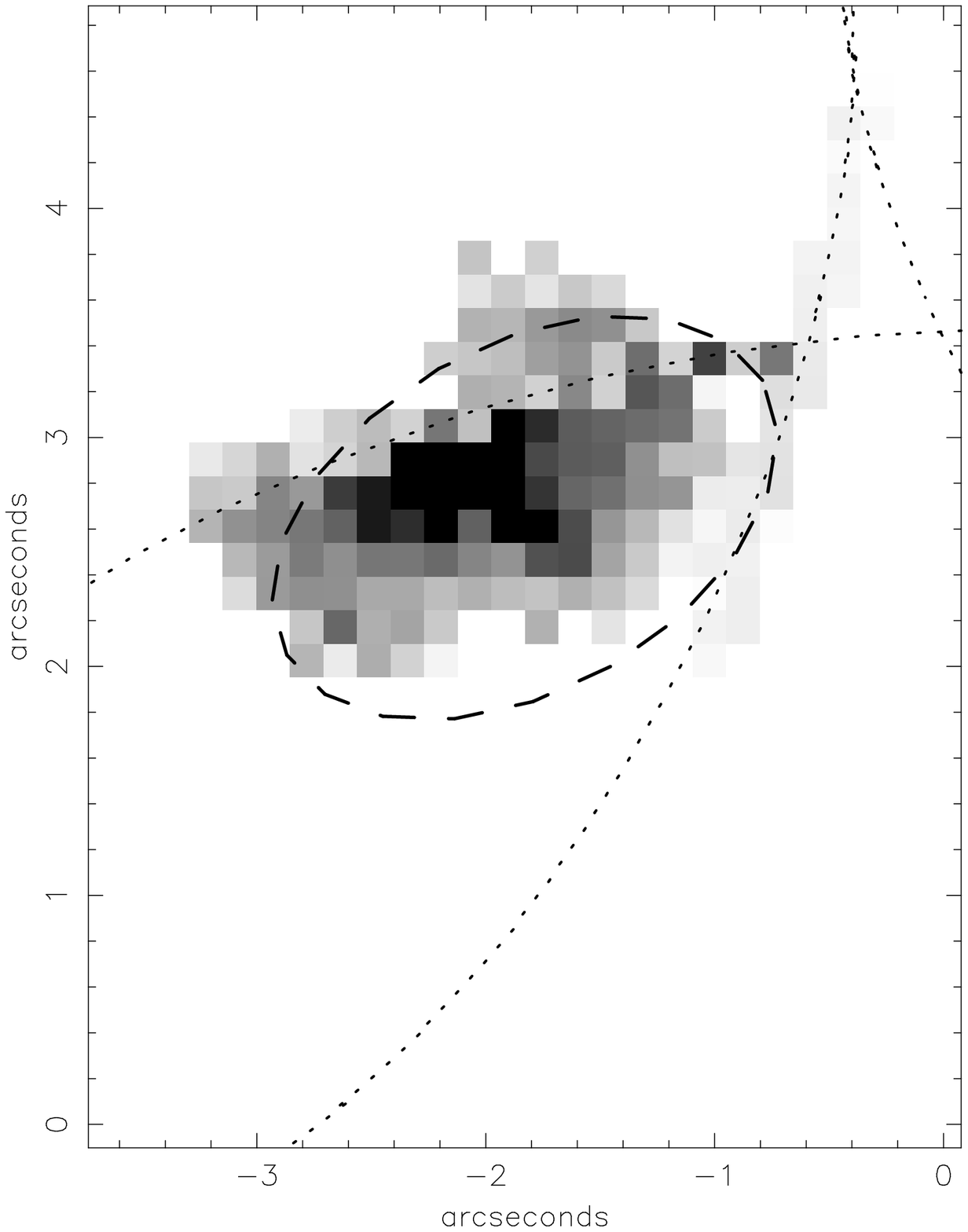}
  \caption{``Cleaned'' source reconstructed behind the PID+cD
    lens (left) and the NFW+cD lens (right). \label{fig:1455S1clean}}
\end{figure}

The collections of isolated pixels in the lower half of the plots are
due to shortcomings of the model.  We remove these spurious pixels by
comparing each datum to its immediate neighbourhood, and discarding data
straying farther than 3 standard deviations from the local average.  The
implications of this ``cleaning'' process are addressed below.  The
``cleaned'' sources are shown in Figure~\ref{fig:1455S1clean}.  The
spurious pixels are gone, but the faint limb responsible for the
tangential arc A3, a feature supported by lensing within the context of
the model, remains.

\placefigure{fig:1455S1clean}

\subsection{Reconstruction of MS~1455}

\begin{figure}[ph]
  \plottwo{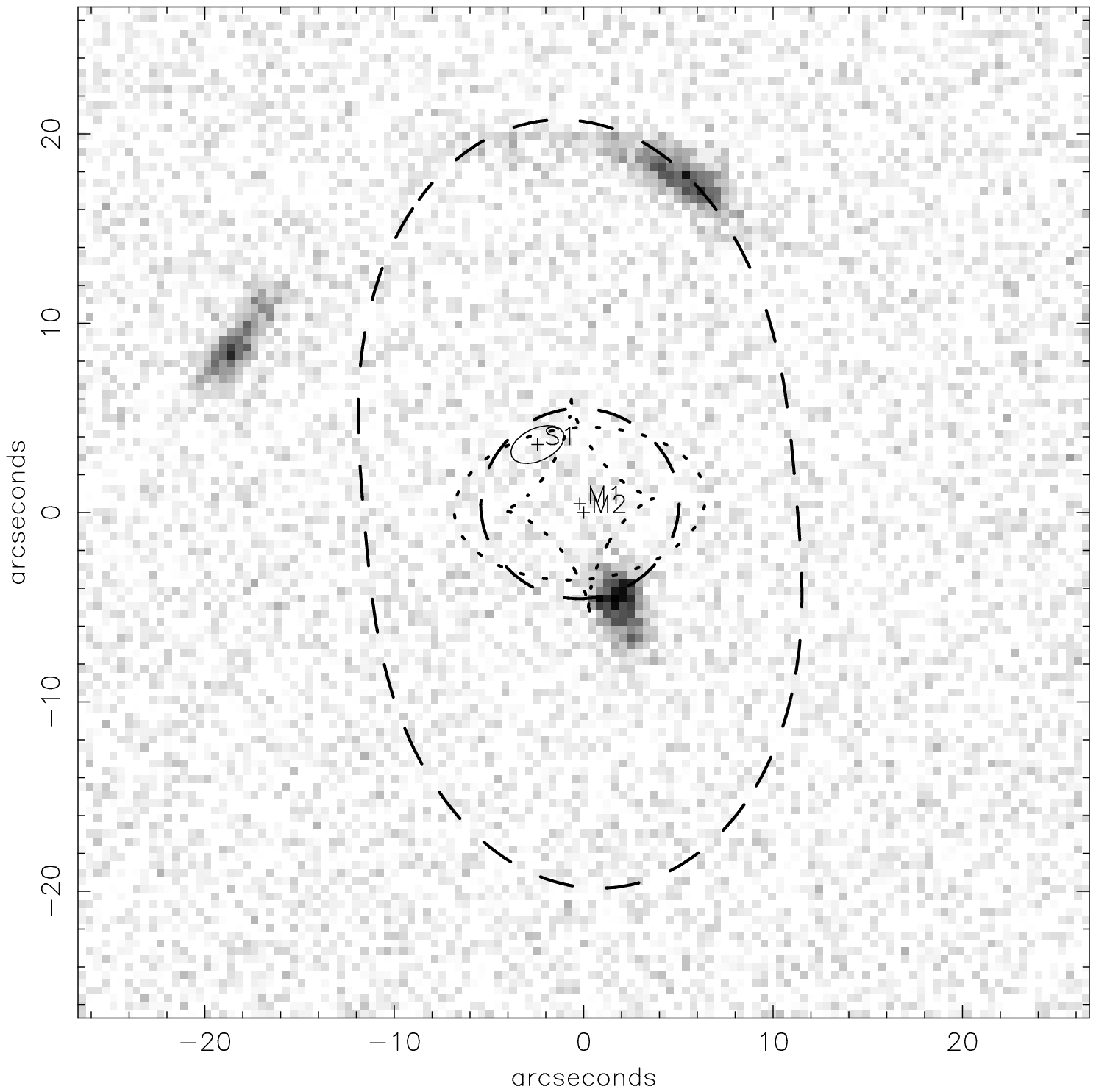}{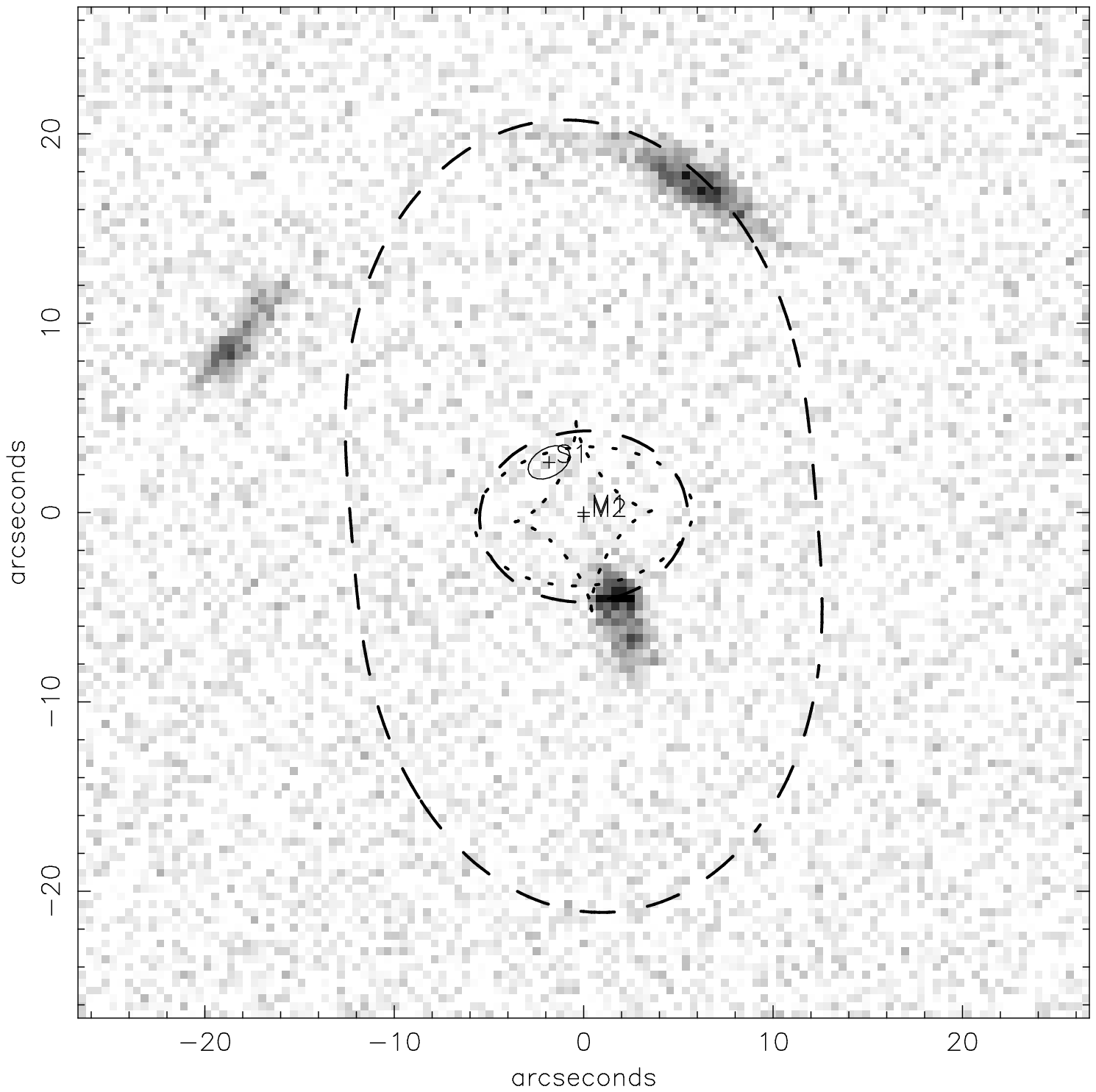}
  \caption{The relensed reconstructed sources in the PID+cD (left) and
    NFW+cD (right) models of MS~1455.  In both cases, the radial arc
    does not extend far enough towards the centre of the lens, and the
    third arc is greatly extended. \label{fig:1455Mxmodels}}
\end{figure}

According to our model of the lens and the source recovered from data,
the lensed features in the observations should be reproduced by passing
the reconstructed source back through the lens model.  The results of
this consistency check are shown in Figure~\ref{fig:1455Mxmodels}.  The
simulations have been convolved against a Moffat point-spread function
with $\beta=2.5$ and radius $R=0\farcs 414$ (two pixels in the
observations) corresponding to a FWHM of $0\farcs 828$ (four pixels),
recreating the seeing at the CFHT at the time of the observations.
Gaussian noise matched to empty regions in the data has been added.
There are two inconsistencies between the observations and the relensed,
reconstructed source: The radial arc does not extend far enough towards
the centre of the lens, and the third arc is much more extended than in
the observations.

\placefigure{fig:1455Mxmodels}

The first flaw can be traced directly to the data ``cleaning'' step.  As
each spurious datum is discarded, its origin in one of the arcs in the
observations is flagged.  These flagged data are concentrated entirely
in the inner end of the radial arc.  By discarding data on the source
plane which originates in only one of three arcs, the reconstructed
source fails to reproduce a portion of the radial arc without effecting
the appearance of the two other arcs.

This behaviour shows that the models we propose for MS~1455 do not
adequately model the mass distribution near the centre of the lens.
Inconsistencies near the centre of the lens are not unexpected, however.
Gravitational lensing does not uniformly measure the mass distribution
of the deflector, but only the cumulative projected mass distribution.
The lensing behaviour far from centre-of-lensing, but still within the
strong regime, is insensitive to perturbations in the mass distribution
at the lens centre.  As demonstrated by \hbox{Miralda-Escud\'{e}} (1995)
in the case of MS~2137, these same perturbations can
remove the radial arc from the lens altogether, because of its proximity
to the centre of the lens.  In MS~1455, the other small galaxies in the
vicinity of the cluster core and the radial arc undoubtedly play a role
in the appearance of the radial arc.  Refinements to the position and
shape of the radial arc can be made by including more masses near the
centre of the cluster.  However, the small number of statistics we
extract from the three lensed arcs does not support the inclusion of
further masses.

Upon relensing the reconstructed source, the third arc A3
appears greatly extended.  This can be traced to the effects of
seeing in the data.  Ideally, the data in any one arc can be traced back
to the source plane, combined to reconstruct the source, and then traced
forward through all three arcs.  The ability of the data from one arc to
reproduce three arcs is a measure of the success of the model.  However,
when the data contained in arc A3 alone is used to reconstruct the
source, only the faint limb of the source following the astroid
tangential caustic, and none of the elliptical bulge, is reconstructed.
When this reconstructed source is passed back through the lens, it forms
a large tangential arc following the critical line of the lens, the
light having originated from the corresponding tangential caustic.

This behaviour indicates that the position of lensed light forming arc
A3 in the observations is not due entirely to gravitational lensing.
Instead, as the simulations is Figure~\ref{fig:1455pidandnfw} suggest, a
bright, very compact arc forms at the location of arc A3.  The image is
smeared due to the effects of seeing.  This broadens the image in the
observations, so that lensing is not wholly responsible for the position
of lensed light in the observations.  To reproduce the enlarged arc, the
inversion algorithm must reconstruct a larger background source, which
is then amplified into a larger relensed arc.  This suggests that the
data should be deconvolved with a suitable point spread function before
applying the inversion algorithm.  Space-based observations of MS~1455
may resolve this problem.

\section{Discussion}

The two-stage inversion algorithm described here decouples the effects
of lens magnification and intrinsic background source structure on the
appearance of the lensed images.  This allows for an independent
reconstruction of the background sources.  That is, the natural
appearance of the background sources is not used to determine the
parametric model.  This is particularly important in the case of
MS~1455, where the faint limb which follows the tangential caustic does
not coincide with the centre of the bright source assumed to be
responsible for all three arcs.  Forcing the pre-images of the three
arcs in MS~1455 to coincide on the source plan leads to inconsistent
lens reconstructions.

The polar moments approach was envisioned to describe galaxy-clusters
with a single centre-of-lensing.  In some clusters, such as A2390
(\cite{PI96.1}), there appear to be more than one centre-of-lensing.
Because of the rapid decrease in deflection with distance from the mass
centre, however, there may be arcs formed primarily about one or the
other centres-of-lensing, which could serve as the origin of the polar
analyses.  Furthermore, relensing the reconstructed sources will test
the consistency of the bimodel mass distribution.  Some suggestion of
this occurs at the centre of MS~1455, where the relensed, reconstructed
radial arc suggests that additional masses are needed.  In the case of
A2390, a parametric lens model built around two centres-of-lensing
should still predict the existence of the ``long straight arc'' formed
by light squeezed between the two centres, even if the arc is not used
to fit the parameters.  Frye \& Broadhurst (1997) suggest that the
``long straight arc'' is in fact a superposition of two arcs, so this
scenario might be moot. 

Further analysis of polar moments is needed, but already they have
several favourable characteristics.  Only four statistics $\bar{r}$,
$\bar{\theta}$, $Q_{rr}$, and $Q_{\theta\theta}$ are required to quite
adequately characterize the position and shape of a wide range of arcs.
The shape parameter $\chi$ is a quantitative measure which can
distinguish between tangential and radial arcs.  Finally, polar moments
may act as a bridge between the strong and weak lensing regimes, for the
dimensions \mbox{$\bar{r}\pm\sqrt{3Q_{rr}}$} and
\mbox{$\bar{\theta}\pm\sqrt{3Q_{\theta\theta}}$} of the annular sector
built from the quadrupole moments smoothly extend into the axes of the
equivalent ellipse which characterizes the shape of a weakly distorted
background galaxy.

We have been unable to define a useful global statistic which can be
minimized to produce a ``best model.''  We could conceive of
constructing a $\chi^2$-like measure which tabulates weighted
differences between the observed and simulated polar statistics.  How
the different statistics $\bar{r}$, $\bar{\theta}$, $Q_{rr}$, and
$Q_{\theta\theta}$ should be weighted, if at all, is unclear.  This
omission is due, in part, to the lack of a quantitative definition of a
``best model'' and a meaningful target value for the $\chi^2$-like
statistic.  At this time, we refrain from forming such a measure, and
rely on the appearance of the relensed, reconstructed sources to check
the consistency of the model.

\section{Conclusions}

We have introduced a two-stage lens inversion algorithm which decouples
the effects of lens magnification and intrinsic source structure in the
appearance of lensed arcs and arclets.  The key to decoupling the
deflector plane from the source plane is characterizing the positions
and shapes of the lensed objects using polar moments.  While these
statistics are artificial, they have interesting and useful qualities.
In reconstructing a background source from demagnified data, we
adopt a simple strategy of generating a pixelized image of the source,
where the resolution of the image is set so that the number of data in
the reconstructed source is comparable to the number of data in the
observations.  Refinements of this strategy, such as adaptive gridding
to take advantage of the concentration of data around the caustics, will
be explored in the future.

It is not the goal of this paper to answer astrophysical questions about
the nature of the cluster-galaxies MS~2137 and MS~1455.  It is
interesting to note, however, that both the non-singular PID and
singular NFW halos are able to model the lensing behaviour of MS~1455.
To begin to answer questions like these requires a detailed model of the
gravitational lens.  The algorithm described here offers an efficient
and intuitive approach to generating such a model.




\newpage

\clearpage
\begin{deluxetable}{lcccccc}
\tablecaption{Polar Moments of Observed MS~2137 Arcs}
\tablehead{%
  \colhead{Arc} &
  \colhead{$\bar{r}$} &
  \colhead{$\bar{\theta}$} &  
  \colhead{$Q_{rr}$} & 
  \colhead{$Q_{r\theta}$} &
  \colhead{$Q_{\theta\theta}$} & 
  \colhead{$\chi$} \\
  \colhead{} & 
  \colhead{(arcsec)} & 
  \colhead{(degrees)}
  } 
\startdata
A0 & $15.40\pm 0.05$ & $-46.7\pm 0.2$ &
$0.15$ & $-2.1$ & $222.7$ & $10.4$ \nl
A2 & $13.07\pm 0.05$ & $-152.1\pm 0.2$ &
$0.37$ & $0.2$ & $11.9$ & $1.3$ \nl
A4 & $19.45\pm 0.05$ & $82.6\pm 0.1$ &
$0.09$ & $-0.5$ & $19.9$ & $5.0$ \nl
AR & $5.43\pm 0.05$ & $-53.6\pm 0.5$ &
$1.07$ & $1.5$ & $5.7$ & $0.2$ \nl
A6 & $24.20\pm 0.05$ & $112.5\pm 0.1$ &
$0.19$ & $0.4$ & $3.4$ & $1.8$ \nl
\enddata
\label{tbl:2137datamoments}
\end{deluxetable}

\begin{deluxetable}{lcccccc}
\scriptsize
\tablecaption{Model Parameters for Simulations of MS~2137}
\tablehead{%
  \colhead{} &
  \multicolumn{2}{c}{Model} &
  \multicolumn{2}{c}{M93} &
  \multicolumn{2}{c}{H97}}
\startdata
Deflector redshift $z_d$ & 
  \multicolumn{2}{c}{$0.313$} &
  \multicolumn{2}{c}{$0.313$} &
  \multicolumn{2}{c}{$0.313$} \nl
Source redshift $z_s$ &
  \multicolumn{2}{c}{$1.05$} &
  \multicolumn{2}{c}{$0.5-3$} &
  \multicolumn{2}{c}{$0.99-1.01$} \nl \nl
Mass distribution & PID & cD & PID & cD & 
  \multicolumn{2}{c}{$\beta$-profile} \nl\tableline
Centre (arcsec) & 
  $(0.00,0.00)$ & 
  $(0.00,0.00)$ & 
  $(0,0)$ & $(0,0)$ & 
  \multicolumn{2}{c}{$(0,0)$} \nl
Orientation (degrees)\,\tablenotemark{a} &
  $6.0$ & $10.0$ &
  $10.5^{+5}_{-9}$\,\tablenotemark{b} & $10.5\pm 5$\,\tablenotemark{b} &
  \multicolumn{2}{c}{$4-10$\, \tablenotemark{c}}  \nl
Eccentricity &
  $0.65$ & $0.45$ &
  $0.62^{+0.08}_{-0.11}$\,\tablenotemark{d} &
  $0.59^{+0.11}_{-0.18}$\,\tablenotemark{d} &
  \multicolumn{2}{c}{$0.583\pm 0.025$\,\tablenotemark{e}} \nl
Mass parameter $\sigma$ &
  $1150$ & $500$ & 
  \nodata & \nodata &
  \multicolumn{2}{c}{\nodata} \nl
$\sigma_{los}$ (km\,s$^{-\!1}$) &
  $1135$ & $676$ &
  $1000$ & $350$\,\tablenotemark{f} &
  \multicolumn{2}{c}{$1216$} \nl
Scale &
  $r_c=4\farcs 0$ & $r_c=0\farcs 51$ &
  $r_c=8\farcs 0^{+0.5}_{-2.0}$ & $1\farcs 5\pm 1$ &
  \multicolumn{2}{c}{$\beta=0.875\pm0.045$}  \nl
   &  & $r_h=1\farcs 50$ & & & 
   \multicolumn{2}{c}{$r_c=2\farcs25\pm 1\farcs75$} \nl \nl
Source model &
  S1 & S2 &
  S1 & S2 &
  S1 & S2 \nl\tableline
Centre (arcsec) &
  $(0.77,1.78)$ & $(-2.71,5.63)$ &
  $(1.0,1.6)$\,\tablenotemark{b} & $(-2.0,4.7)$\,\tablenotemark{b} &
  \multicolumn{2}{c}{$3.7\pm 0.3$ separation} \nl
Semi-major axis (arcsec) &
  $0.65$ & $0.65$ &
  \nodata & \nodata &
  $0.58\pm 0.025$ & \nodata \nl
Orientation (degrees)\,\tablenotemark{a} &
  $25.0$ & $9.0$ &
  \nodata & \nodata &
  \nodata & \nodata \nl
Eccentricity &
  $0.74$ & $0.74$ &
  \nodata & \nodata &
  $0.81$\,\tablenotemark{e} & \nodata \nl
\enddata
\tablenotetext{a}{Counter-clockwise from positive $x$-axis.}
\tablenotetext{b}{Estimated.}
\tablenotetext{c}{Estimated, based on \emph{HST} orientation.}
\tablenotetext{d}{Deduced from ellipticity $(a^2-b^2)/(a^2+b^2)$.}
\tablenotetext{e}{Deduced from ellipticity $1-b/a$.}
\tablenotetext{f}{Derived from Faber-Jackson relation.}
\label{tbl:2137paramtbl}
\end{deluxetable}

\begin{deluxetable}{lcccccc}
\tablecaption{Polar Moments of Simulated MS~2137 Arcs}
\tablehead{%
  \colhead{Arc} &
  \colhead{$\bar{r}$} &
  \colhead{$\bar{\theta}$} &  
  \colhead{$Q_{rr}$} & 
  \colhead{$Q_{r\theta}$} &
  \colhead{$Q_{\theta\theta}$} & 
  \colhead{$\chi$} \\
  \colhead{} & 
  \colhead{(arcsec)} & 
  \colhead{(degrees)}
  } 
\startdata
A0 & $15.5\pm 0.2$ & $-44.7\pm 0.7$ &
$0.1$ & $-0.3$ & $177.0$ & $11.4$ \nl
A2 & $13.3\pm 0.2$ & $-151.8\pm 0.9$ &
$0.2$ & $0.1$ & $10.6$ & $1.7$ \nl
A4 & $19.4\pm 0.2$ & $84.9\pm 0.6$ &
$0.1$ & $-0.3$ & $11.9$ & $3.7$ \nl
AR & $6.2\pm 0.2$ & $-54.5\pm 1.8$ &
$1.5$ & $0.6$ & $15.1$ & $0.3$ \nl
A6 & $24.3\pm 0.2$ & $109.2\pm 0.5$ &
$0.1$ & $0.1$ & $4.1$ & $2.7 $ \nl
\enddata
\label{tbl:2137modelmoments}
\end{deluxetable}

\begin{deluxetable}{lcccccc}
\tablecaption{Polar Moments of Observed MS~1455 Arcs}
\tablehead{%
  \colhead{Arc} &
  \colhead{$\bar{r}$} &
  \colhead{$\bar{\theta}$} &  
  \colhead{$Q_{rr}$} & 
  \colhead{$Q_{r\theta}$} &
  \colhead{$Q_{\theta\theta}$} & 
  \colhead{$\chi$} \\
  \colhead{} & 
  \colhead{(arcsec)} & 
  \colhead{(degrees)}
  } 
\startdata
A1 & $5.1\pm 0.1$ & $-70.1\pm 1.2$ &
$3.9$ & $-3.9$ & $61.9$ & $0.4$ \nl
A2 & $20.5\pm 0.1$ & $154.8\pm 0.3$ &
$0.4$ & $1.8$ & $29.8$ & $3.1$ \nl
A3 & $18.3\pm 0.1$ & $71.7\pm 0.3$ &
$0.7$ & $1.1$ & $6.7$ & $1.0$ \nl
\enddata
\label{tbl:1455datamoments}
\end{deluxetable}

\begin{deluxetable}{lcccc}
\tablecaption{Model Parameters for Simulations of MS~1455}
\tablehead{%
  \colhead{} &
  \multicolumn{2}{c}{PID+cD} &
  \multicolumn{2}{c}{NFW+cD}}
\startdata
Deflector redshift $z_d$ & 
  \multicolumn{2}{c}{$0.257$} &
  \multicolumn{2}{c}{$0.257$} \nl
Source redshift $z_s$ &
  \multicolumn{2}{c}{$0.825$} &
  \multicolumn{2}{c}{$0.620$} \nl \nl
Mass distribution & PID & cD & NFW & cD \nl\hline
Centre (arcsec) & 
  $(-0.2,0.5)$ &   $(0.0,0.0)$ & 
  $(0.0,-0.2)$ & $(0,0)$ \nl
Orientation (degrees)\,\tablenotemark{a} &
  $-86.0$ & $-56.0$ &
  $-86.0$ & $-56.0$ \nl
Eccentricity &
  $0.75$ & $0.80$ &
  $0.70$ & $0.80$ \nl
Mass parameter $\sigma$ &
  $1143$ & $500$ & 
  $2734$ & $470$ \nl
$\sigma_{los}$ (km\,s$^{-\!1}$) &
  $1133$ & $687$ &
  $1133$ & $729$ \nl
Scale &
  $r_c=4\farcs 0$ & $r_c=0\farcs 57$ &
  $r_s=40\farcs 0$ & $r_c=0\farcs57$ \nl
  & $r_h=1\farcs50$ &
  & $r_h=1\farcs50$ \nl\nl
Source model &
  \multicolumn{2}{c}{S1} &
  \multicolumn{2}{c}{S1} \nl\hline
Centre (arcsec) &
  \multicolumn{2}{c}{$(-2\farcs 44,3\farcs 59)$} &
  \multicolumn{2}{c}{$(-1\farcs 83,2\farcs 65)$} \nl
Semi-major axis (arcsec) &
  \multicolumn{2}{c}{$1.50$} &
  \multicolumn{2}{c}{$1.20$} \nl
Orientation (degrees)\,\tablenotemark{a} &
  \multicolumn{2}{c}{$25.0$} &
  \multicolumn{2}{c}{$30.0$} \nl
Eccentricity &
  \multicolumn{2}{c}{$0.82$} &
  \multicolumn{2}{c}{$0.78$} \nl
\enddata
\tablenotetext{a}{Counter-clockwise from positive $x$-axis.}
\label{tbl:1455paramtbl}
\end{deluxetable}

\begin{deluxetable}{lcccccc}
\tablecaption{Polar Moments of Simulated MS~1455 Arcs}
\tablehead{%
  \colhead{Arc} &
  \colhead{$\bar{r}$} &
  \colhead{$\bar{\theta}$} &  
  \colhead{$Q_{rr}$} & 
  \colhead{$Q_{r\theta}$} &
  \colhead{$Q_{\theta\theta}$} & 
  \colhead{$\chi$} \\
  \colhead{} & 
  \colhead{(arcsec)} & 
  \colhead{(degrees)}
  } 
\startdata
PID+cD \nl
A1 & $5.7\pm 0.2$ & $-69.4\pm 2.1$ &
  $3.9$ & $-2.5$ & $49.4$ & $0.4$ \nl
A2 & $20.2\pm 0.2$ & $153.6\pm 0.4$ &
  $0.4$ & $2.0$ & $38.5$ & $3.5$ \nl
A3 & $17.8\pm 0.2$ & $70.1\pm 0.7$ &
  $0.1$ & $0.1$ & $5.1$ & $2.2$ \nl\nl
NFW+cD \nl
A1 & $6.4\pm 0.2$ & $-70.8\pm 1.9$ &
  $4.8$ & $-1.4$ & $31.6$ & $0.3$ \nl
A2 & $20.3\pm 0.2$ & $154.5\pm 0.6$ &
  $0.3$ & $1.8$ & $37.6$ & $4.0$ \nl
A3 & $17.9\pm 0.2$ & $67.4\pm 0.7$ &
  $0.1$ & $0.1$ & $11.6$ & $3.4$
\enddata
\label{tbl:1455modelmoments}
\end{deluxetable}

\end{document}